\documentclass{svjour2}                    
\smartqed  
\usepackage{graphicx}
\usepackage{amssymb,amsmath}
\usepackage{changes}
\usepackage{physics,gensymb}
\usepackage{booktabs}

\begin{document}

\title{Dimension reduction for systems with slow relaxation. \thanks{This work was funded in part by a grant from GoMRI. We also received support from NSF-DMS-1109856 and NSF-OCE-1434198. }
}
\subtitle{In memory of Leo P. Kadanoff}

\titlerunning{Slow Relaxation and Dimension Reduction}        

\author{Shankar C. Venkataramani  \and
        Raman C. Venkataramani \and
        Juan M. Restrepo 
}

\authorrunning{S.C. Venkataramani, R.C. Venkataramani, J.M. Restrepo} 

\institute{Shankar C. Venkataramani \at
              Department of Mathematics,  
              University of Arizona, Tucson, AZ 85721 \\
              \email{shankar@math.arizona.edu}        
           \and
           Raman C. Venkataramani \at
           Seagate Technology, 389 Disc Dr., Longmont, CO 80503 \\
           \email{ramanv@ieee.org} 
           \and 
           Juan M. Restrepo \at
           Department of Mathematics
           Oregon State University, Corvallis OR, 97330    \\
           \email{restrepo@math.oregonstate.edu}        
}

\date{Received: date / Accepted: date}

\maketitle

\begin{abstract}  
 We develop reduced, stochastic models for high dimensional, dissipative dynamical systems that relax very slowly to equilibrium and can encode long term memory. We present a variety of empirical and first principles approaches for model reduction, and build a mathematical framework for analyzing the reduced models. We introduce the notions of universal and asymptotic filters to characterize `optimal' model reductions for sloppy linear models. We illustrate our methods by applying them to the practically important problem of modeling evaporation in oil spills.

\keywords{dimension reduction \and slow relaxation \and sloppy models \and Mori-Zwanzig projection \and multi-scale \and aging \and weathering \and glassy systems \and oil spills}
 \PACS{05.40.-a \and 05.10.Gg \and 89.75.Da \and 82.20.Db \and 92.20.Ny}
\end{abstract}

\normalem

\section*{In Memoriam}

{\em It is with immense gratitude that we dedicate this article to Leo Kadanoff. Two of the authors (SV and JR) first met Leo as postdocs. Our lives would have been very different if not for the outsize role that Leo played in our professional development and also in our personal growth. His door, and his mind, were always open. 
He reminded us to  ask questions,  about how  humility gave the courage to know what we knew and we  did not. He taught us to be fearless about pursuing a wide range of interests. The fun he had with science was infectious, and the skill he had to ask the right questions is impossible to match.

Our experiences were by no means unique. There are hundreds of people whose lives Leo touched in the same way.  So many of his  informal seminar or lunch questions turned  into  
full  research enterprises. It is no exaggeration to say that  at some point people needed only to know that Leo had been the one to ask the question in order to assure themselves that their scientific investigations were worthwhile. 

The last time that one of us saw Leo was in May of 2015. Coincidentally, it was at a talk  on the subject of oil spill modeling, and included some of the rudimentary ideas that grew into this paper. Leo came down to the (new) James Franck Institute for the talk. He was just as sharp as ever, and he made sure that the graduate students in the audience got all the physical intuition that the speaker elided over, by interjecting appropriately. It was classic Leo. How we miss him! }

\tableofcontents 

 \section{Introduction}
   
 The methods of equilibrium statistical physics are remarkably successful in characterizing the thermodynamic limit of Hamiltonian systems with many interacting degrees of freedom. Considering adiabatic perturbations of the equilibrium state, one can also compute the transport coefficients that characterize the linear response to external forcing in non-equilibrium states. 
 
 For many out-of-equilibrium systems, the relaxation to equilibrium is exponential and is governed by linear response theory. There are interesting examples that do not fit this paradigm -- {\em Glassy systems}  relax very slowly (typically logarithmically) \cite{Bouchaud_Aging_1999}. Such systems include spin glasses \cite{Bouchaud_Out_1998},  forced crumpling \cite{kittiwit_crumpling} and the stick-slip phenomenon and evolution of frictional strength \cite{Ben-David_Slip_2010}. In these systems, the relaxation to a {\em putative equilibrium} state is extremely slow and it is unclear if an equilibrium can {\em ever be reached}. 

A striking example in this direction is the phenomenon of {\em aging} \cite{Bouchaud_Aging_1999,Amir_On_2012}.  A glassy system is allowed to relax over a long time scale and is then driven by an external perturbation for a length of time $t_w$, the waiting time. The perturbation is then removed, and the system is allowed to relax once again. In this situation, the subsequent relaxation dynamics depends not only on the (initially perturbed) macroscopic state of the system, but also on the waiting time $t_w$, showing that the relaxation of the system depends on details of its micro-state, and that the micro-state has {\em memory}, i.e. it can `record' aspects of the history of the system. In particular the dynamics can distinguish identical macro-states that result from distinct preparations (e.g. different waiting times) of an initially ``relaxed'' or ``equilibrium" sample. 
 
There are other consequences of slow relaxation to equilibrium. In amorphous materials below the glass transition temperature $T_g$, the slow relaxation manifests itself as a slow change in the effective properties of the material, a process called {\em physical aging}. As we discuss below, a similar phenomenon is very important in the modeling of oil spills, although the mechanisms are different and crude oil is not a ``glassy system" in the usual understanding of this term.

The evolution of oil in the environment is also called {\em Weathering}. Weathering  is informally defined as the mechanical or chemical changes that occur in compounds, such as rock, due to exposure. Its distinguishing feature is that it occurs over very long times.
Weathering is a process that is distinct from glassy dynamics, nevertheless, it shares the feature of slow relaxation with the latter. 

Our interest in capturing such weathering dynamics arose from work developing a model for oil spill transport in the ocean \cite{nearshore,oil1,oil2}. Crude oil is made up of hundreds, even thousands of different chemical compounds. Oil  will react chemically  due to its complex chemistry, its their exposure to the elements, and due to biological action \cite{Hoult_Oil_1969,Spaulding_A_1988}.  These changes occur at many time scales (from hours to decades) without any clear scale separation \cite{Fingas_A_1995}. 

Oil is both a major  economic commodity and a particularly dangerous pollutant so there is great urgency in addressing a spill as soon as it happens.   A transport model for an oil spill is thus not very useful if it can only model the asymptotic state. This is because the buoyancy, surface tension, and viscosity of oil droplets depend on their chemical composition and are changing, due to exogenous and endogenous reactions, at different rates at different times. These changes on the micro-scale have dramatic consequences on the large scale dynamics of expansive oil spills, and on how these oil spills are going to be contained. Petroleum and chemical engineers are thus keenly interested in understanding and effectively modeling these changes, and one of the goals in this work is to develop methods to construct reduced dimensional models to this end (See \cite{Spaulding_A_1988,Fingas_A_1995} for a review of existing modeling approaches). 
We seek to develop low dimensional, stochastic models for multi-scale, dissipative dynamics, that can be applied to the practically important problems. 

The other goal of this work is just as important -- we want to develop intuition and new approaches for problems in statistical physics including (i) coarse graining for systems with slow relaxation/long memory and (ii) robust predictions for {\em sloppy models} \cite{Brown_Statistical_2003,Waterfall_Sloppy_2006}. 

 Recently, several techniques have been proposed to address the issue of dimension reduction for complex systems using tools from non-equilibrium statistical mechanics\cite{Chorin_Optimal_2000,Givon_Extracting_2004,Darve_Computing_2009,Stinis_Renormalized_2015,Venturi_Mori_2015}.  These techniques are based on earlier work by Mori \cite{Mori_Transport_1965}, Zwanzig \cite{Zwanzig_Nonlinear_1973,Zwanzig_problems_1980} and Kawasaki \cite{Kawasaki_Simple_1973}. The Mori-Zwanzig projection operator formalism \cite{Zwanzig_Nonequilibrium_2001,Van_Kampen_Stochastic_2007} decomposes the high-dimensional phase space of a system into resolved (or observed) variables and unresolved/unobserved degrees of freedom. The key idea is to project the full dynamics on the phase space (more properly, to project the Liouville equation for the evolution of probability measures on phase space) on to dynamics on the resolved degrees of freedom using statistical information to take an expectation over the unresolved degrees of freedom \cite{Zwanzig_Nonequilibrium_2001}. This procedure results in a {\em generalized Langevin equation} (GLE)
$$
\dot{X} = f(X) + \int_{0}^t K(X(t-s),s) ds + \eta(X,t),
$$
where $X \in \mathbb{R}^n$ is the set of $n$ resolved degrees of freedom, $K(X(t-s),s)$ is the {\em memory kernel}, that quantifies the information in the history of the resolved variables $X$ on its subsequent evolution, and $\eta \in \mathbb{R}^n$ the ``noise" is governed by the {\em orthogonal dynamics} \cite{Givon_Existence_2005,Darve_Computing_2009}.

In situations with an invariant measure on phase space, and one uses a linear Mori projection \cite{Mori_Transport_1965}, the resulting GLE is  linear, and the memory kernel and the noise covariance are related by the Fluctuation-Dissipation theorem \cite{Kubo_The_1966}. If the memory kernel $H$ decays exponentially one can truncate the memory integral. Approximating the memory kernel $H(s)$ in terms of particular families of functions \cite{Zwanzig_Nonequilibrium_2001,Van_Kampen_Stochastic_2007,Venturi_Convolutionless_2014,Kondrashov_Data_2015} gives rise to autonomous, stochastic differential equations that approximate the evolution of the resolved degrees of freedom $z(t)$, and for a particular ``nonlinear" Zwanzig projection which gives the evolution of the conditional expectation of the current state,  i.e  an optimal predictor for the resolved variables \cite{Chorin_Optimal_2000,Chorin_Optimal_2002}. In this work, we will explore to what extent these methods carry over to situations without a non-trivial invariant measure on phase space.

This paper is organized as follows. In Sec.~\ref{sec:model1} we review some of the literature on modeling the evaporation of crude oil, and present a simplified model for this process. In Sec.~\ref{sec:mz} we review the Mori-Zwanzig projection operator formalism focusing on the discrete time setting as in Darve et al \cite{Darve_Computing_2009}. Section~\ref{sec:parametrization} reviews some basic ideas in signal processing and then discusses various approaches to dimension reduction/stochastic modeling using autonomous/shift-invariant filters. In Sec.~\ref{sec:analysis} we develop an analytic framework that allows us to design non-autonomous/time-varying reduced models for systems with slow relaxation. In Sec.~\ref{sec:MSM} we apply our methods to weathering in oil spills and present a concluding discussion in Sec.~\ref{sec:discuss}.

\section{A Dynamical System with Slow Relaxation}
\label{sec:model1}

There is considerable interest in the evaporation process of crude oil, since this is an important process in the initial stages of an oil spill. Typical crude oil spills in the ocean can lose up to 40\% or more of their oil to evaporation in the first few days before other important processes, e.g. the emulsification of oil in water, have a significant effect \cite{Fingas_modeling_2013}.

Theoretical approaches to modeling oil evaporation \cite{Mackay_Evaporation_1973,Stiver_Evaporation_1984,Fingas_Modeling_2004} are based on modifying equations for the (much better understood process of) evaporation of water \cite{Sutton_Wind_1934}. Thermodynamics governs the process by which water molecules from the liquid enter the vapor phase at an air-water interface. However, this is not the limiting factor which determines the rate of evaporation. Rather, the rate of evaporation is regulated by the saturation of the air boundary-layer near the water surface. Indeed, dry air can hold up to a certain (temperature dependent) maximum amount of water vapor (the relative humidity cannot be more than 100\%), and once the boundary layer reaches this level of saturation, the rate of evaporation is essentially governed by how quickly the water vapor can be transported (turbulent diffusion, laminar flow, etc) away from the water surface. Using similar ideas, the evaporation of each compound in oil can be modeled as 
$$
E/C \approx K T S
$$
where $E$ is the evaporation rate, $C$ the concentration of the compound, $K$ is a mass transfer rate, $T$ is a coefficient that characterizes the turbulent/laminar transport of the vapor away from the interface and $S$ is a factor that depends on the saturation of the boundary layer by the evaporating fluid  (cf. Eq.~(1) in \cite{Fingas_modeling_2013}). The quantity $\alpha = KTS$ is the evaporation rate constant, and various theoretical/empirical approaches to for obtaining expressions for $T$ and $S$ are discussed in detail by Fingas \cite{Fingas_modeling_2013}.

Based on these considerations, we investigate a simple, generic dynamical system that exhibits slow relaxation as a model for the evaporation of  `oil', thought of as a composite with many individual species. This system is solvable and thus offers a benchmark for measuring the performance of reduced dimensional models. 
  
We will assume that `oil' consists of $I$ distinct species with concentrations $c_i(t), i = 1,2,\ldots,I$ each decaying  at a constant rate $\alpha_i = K_i T S_i$, to obtain
\begin{equation}
     \partial_t c_i(t) = - \alpha_i c_i(t), \qquad \alpha_i > 0, 1 \leq i \leq I.
      \label{eq:species}
      \end{equation}
      We can also think of this system as describing non-interacting eigenmodes in a dissipative system that is relaxing to its equilibrium. $I \gg 1$ will be assumed  very large. It is in this situation that a low-dimensional approximation of the system is particularly useful. We can chose the indexing so that $\alpha_{i+1} > \alpha_i$ (identifying ``chemically distinct" species with the same evaporation rate as a single virtual species). The species with $\alpha$ close to zero decay very slowly, so this model is similar in spirit to the approach of Amir et al \cite{Amir_On_2012} for  studying aging in glasses. 
            
       Note that, every concentration $c_i$ is decaying in time, so the eventual state is one with no oil -- the invariant measure is a singular measure corresponding to a point mass at $0$. This system is dissipative, so the methods of equilibrium statistical mechanics of Hamiltonian systems do not directly apply. The projection operator formalism, however,  {\em does not rely} on the underlying dynamics on phase space being Hamiltonian, and the Mori-Zwanzig approach is thus applicable  \cite{hald}.
      
  We consider the case of a single source of measurements (i.e. a scalar function $M(t)$).  We assume that the observation $M(t)$ is a weighted average of the concentrations $c_i$:
      $$M(t) = \sum_i \beta_i c_i(t) = \sum_i \beta_i c_i(0) e^{-\alpha_i t}.
      $$ 
      It is impractical/impossible to separately measure the concentrations/amounts $c_i$ of {\em all} the individual species.   One approach that naturally suggests itself is to use the measured quantity $M(t)$ to extract the various decay rates $\alpha_i$ using nonlinear fitting. {\em This approach will not work} \cite{Waterfall_Sloppy_2006}. Indeed, the discrete oil evaporation model~\eqref{eq:species} is a prototypical example of a sloppy model \cite{Waterfall_Sloppy_2006}. It is identical to the model for the mixture of radioactive nuclides considered in Waterfall {\em et al} \cite{Waterfall_Sloppy_2006}, and as they show,  one cannot hope to extract the decay rates $\alpha_i, i = 1,2,\ldots,I$ from the measured function $M(t)$ \cite{Waterfall_Sloppy_2006,Transtrum_Geometry_2011}.
      
     We will therefore consider the complementary limit, where the number of distinct species   distinct species $I \gg 1$. We will not attempt to identify the individual species in the mixture; rather we will use ideas from signal processing and statistical mechanics to make predictions for quantities that are robust \cite{Brown_Statistical_2003} and insensitive to the precise composition of the oil. 
     To this end,  we will assume that $\beta_i$ and $\alpha_i$ depend ``smoothly" on $i$, i.e we can interpolate the values of $\alpha_i$ and $\beta_i$ to get functions that only vary on scales $\Delta i \gg 1$. Thus, we can replace the discrete index $i$ by a continuous variable $w$ 
      \begin{equation}
      \alpha_i = \alpha_{min} (1-w) + \alpha_{max} w,
      \label{eq:decay-rates}
      \end{equation} where $\alpha_{min} = \alpha_1$ is the evaporation rate of the least volatile species and $\alpha_{max} = \alpha_I$ is the evaporation rate of the most volatile species. A natural time scale for the system is 
      \begin{equation}
      t_0 = \frac{1}{\alpha_{max} - \alpha_{min}},
      \label{eq:tscl}
       \end{equation}
       corresponding to the time by which the relative concentration 
       $$\frac{c_I(t_0)}{c_1(t_0)} = \frac{1}{e} \frac{c_I(0)}{c_1(0)}. 
       $$
     We will now obtain the equations for the continuum limit $I \to \infty$ for time scales $t \gtrsim t_0$ i.e. on time scales on which the relative concentrations of the various species vary significantly. Let $i(w)$ connote a smooth, monotonic interpolation of the inverse of the function 
      $$
      w(i) = \frac{\alpha_i-\alpha_{min}}{\alpha_{max} - \alpha_{min}},
      $$
      and $\rho(w,t)$ the  smooth interpolation of the function
      $$
      \rho(w(i),t) = e^{\alpha_{min} t} \beta_i c_i(t) \cdot \frac{\alpha_{max}-\alpha_{min}}{\alpha_{i+1} - \alpha_i}.
      $$
      A direct calculation now yields
      \begin{align*}
      M(t) & = \sum \beta_i c_i(t) \\
      & = e^{-\alpha_{min} t} \sum \rho(w(i),t) [w(i+1)-w(i)].
      \end{align*}
      Using the differential equation for $c_i$, we get 
      $$ \partial_t \rho(w(i),t) = -(\alpha_{min}+ (\alpha_{max} - \alpha_{min})w(i)) \rho(w(i),t).
      $$ 
      We rescale to a non-dimensional time $\tilde{t} = \frac{t}{t_0} = (\alpha_{max} - \alpha_{min}) t$ and consider the case $\alpha_{min} =0$ to describe slow relaxation. Indeed, for crude oil in the environment, $\alpha_{min} \approx 0$ and it is in the scale of 1/decade.  
      
      Taking the (naive) continuum limit in the two previous equations, and dropping the tildes on $\tilde{t}$ with the understanding that henceforth $t$ is dimensionless, we obtain
      \begin{equation}
      \partial_t \rho(w,t) = -w \rho(w,t), \qquad M(t) = \int_0^1 \rho(w,t) \, dw.
      \label{eq:continuum}
      \end{equation}
      We will refer to this equation, along with an appropriate {\em random} initial condition $\rho(w,0)$ as the {\em linear evaporation process}. The only assumptions that went into the derivation of the continuum limit are,  (1) $I \gg 1$; (2) $\alpha_i$ and $\beta_i$ vary on scales $\Delta i \gg 1$. The continuum limit equations are {\em independent} of the precise details of $\alpha_i$ and $\beta_i$, which  are absorbed into the change of variable from $c_i(t)$ to $\rho(w,t)$.  
      
      For a complete specification of the problem, we need to characterize the statistics of the initial measure $\rho(w,0)$, which is random, reflecting the uncertainties in the initial composition of the oil. In particular, the statistics of $\rho(w,0)$ should be inferred from the statistics of the concentrations $c_i(0)$ (of the discrete species), but one needs some attention to  manner in which we take the limit $I \to \infty$ so that we do not get a deterministic (instead of random) limit for $\rho(w,0)$. This would indeed be the case, from the central limit theorem, if the various $c_i(0)$ were i.i.d random variables with finite variance and distributions independent of $I$. This issue is somewhat subtle, and address it in  Section~\ref{sec:analysis}.

We can explicitly solve~\eqref{eq:continuum} to get $
\rho(w,t) = \rho(w,0) e^{-wt}.
$
This is the Schr{\"o}dinger picture of the evolution of the system, in which the measure associated with the state of the oil changes with time. 

The Heisenberg picture arises from considering {\em linear observables}  given by pairing the measure $d \mu^t \equiv \rho(\omega,t) d \omega$ with (an appropriate subset of) continuous functions on $[0,1]$. Every continuous function $g$ is associated with a linear observable $G$ given by
$$
G(t) = \int g(w) \rho(w,t) dw.
$$ 
The total mass $M(t)$ is thus the observable associated with the constant function $f(x) = 1$. In the Heisenberg picture we have
\begin{equation}
G(t) = \int g(w) \rho(w,t) dw  = \int g(w) e^{-wt} \rho(w,0) dw = \int g^t(w) d\mu^0(w).
\label{heisenberg}
\end{equation}
The evolution of the continuous function associated with an observable $G$ is gven by $g^t(w) = g(w) e^{-wt}$ so that $\partial_t g^t(w) = - w g^t(w)$.

We note the contrast between our system~\eqref{eq:continuum}, and the typical situation of a dynamical system $\dot{x} = f(x)$ on a  high-dimensional phase space $\Sigma$. For the dynamical system, the Heisenberg picture is given by evolving continuous functions on $\Sigma$ through
$
g^t(x) = g(\varphi(x,t))
$
where $\varphi$ is the solution map, i.e $\varphi(x,0) = x$ and $\partial_t \varphi(x,t) = f(\varphi(x,t))$. In particular, the constant function 1 is an invariant under this evolution. $g(x) = 1$ for all $x$ in $\Sigma$ implies that $g^t(x) =1$ for all $x$ and $t$. In contrast, the  evolution in \eqref{heisenberg} which gives $g^t(w) = e^{-wt}$. Consequently, the continuum limit \eqref{eq:continuum} {\em is not the Liouville equation} for a dynamical system. 

We will henceforth work in a discrete time setting, that can be viewed as a Takens delay-coordinate embedding \cite{takens} of the continuous time system. There are many reasons to do this, including the difficulty in parametrizing continuous time stochastic processes \cite{Chorin_Discrete_2015,Lu_Data_2017} and the fact that, for our application to oil spills, the sensor data is only obtained at discrete time intervals. We can recast \eqref{eq:continuum}~and~\eqref{heisenberg} as maps in discrete time by defining $t = n \tau, \rho_{n}(w) \equiv \rho(w,n\tau)$. 
With these substitutions, 
\begin{align} 
\rho_{n+1}(w) & = \Lambda^T \rho_{n}(w), \label{Xfr-Koopman} \\
g^{(n+1)\tau}(w) & = \Lambda g^{n \tau} (w), \nonumber
\end{align}
where $\Lambda$ is a bounded operator on $C([0,1])$ that takes a continuous function $g(w)$ to $\Lambda g(w) \equiv e^{-w\tau} g(w)$, and $\Lambda^T$ is the adjoint on the dual space of measures on $[0,1]$. Note that $\Lambda$ extends naturally to a self-adjoint operator on $L^2([0,1])$ also defined by $\Lambda h(w) = e^{-w\tau} h(w)$ for all $h \in L^2$, so we can also consider the (larger) set of oservables given by $L^2$ functions on $[0,1]$.
In this  case the density $\rho$ is also in $L^2$ and the evolution is given by a self-adjoint operator $\Lambda^T = \Lambda$ on $L^2([0,1])$. Although the maps in~\eqref{Xfr-Koopman} are not the transfer operator \cite{baladi-book} (respectively the Koopman operator \cite{Budisic_Applied_2012}) corresponding to a dynamical system, they have the same formal structure so we will attempt to use discrete time projection operator techniques for model reduction \cite{Chorin_Discrete_2015,Lu_Data_2017,Lin_Stochastic_2016}.
      
      \section{The Mori-Zwanzig projection formalism} \label{sec:mz}
      
    We first present a short review of the discrete-time Mori-Zwanzig projection formalism following the presentation in Darve et al \cite{Darve_Computing_2009}.  The setup is as follows: $\mathcal{H}$ is a Hilbert space and $\Lambda: \mathcal{H} \to \mathcal{H}$ is a linear operator on this space. We can think of $\Lambda$ as $e^{\tau \mathcal{L}}$ where $\tau$ is a `time-step' and  $\mathcal{L}$ is the Liouville operator (the generator) evolving measures on phase space $\Sigma$ and $\mathcal{H} \subseteq \mathcal{M}(\Sigma)$ so every element of $\mathcal{H}$ can be interpreted as a (signed) measure on $\Sigma$ . Linear observables are given by linear operators $g: \mathcal{H} \to \mathbb{R}$, so the set of linear observables is the dual $\mathcal{H}^* = \mathcal{H}$. A (general, nonlinear) observable is any (measurable) function of a finite collection of linear observables, so the observables form an algebra of mappings $\mathcal{H} \to \mathbb{R}$. Finally, $P: \mathcal{H} \to \mathcal{H}$ is an orthogonal projection and $Q = I - P$ is the complementary projection. We will use the bra-ket notation and represent states (elements of $\mathcal{H}$) by ket-vectors and linear observables by bra-vectors.
    
    We consider the discrete time dynamical system $\ket{\rho_{n+1}} = \Lambda \ket{\rho_n}$. We decompose $\ket{\rho_n} = \ket{\xi_n} + \ket{\eta_n}$ where $\ket{\xi_n} = P \ket{\rho_n}$ (the observations) and $\ket{\eta_n} = Q \ket{\rho_n}$. An elementary argument by induction shows that 
    \begin{align}
    \label{mz_discrete}
   \ket{ \rho_{n}} & = \ket{\xi_n} + \ket{\eta_n} \\
    & = \ket{\xi_n} + Q \Lambda (\ket{\xi_{n-1}} + \ket{\eta_{n-1}}) \nonumber \\
 & = \ket{\xi_n} + Q \Lambda \ket{\xi_{n-1}} + (Q \Lambda)^2 (\ket{\xi_{n-2}} + \ket{\eta_{n-2}}) \nonumber \\
 & = \ket{\xi_n} + Q \Lambda \ket{\xi_{n-1}} + (Q \Lambda)^2 \ket{ \xi_{n-2}} + \cdots + (Q \Lambda)^n \ket{\xi_{0}} + (Q \Lambda)^n Q \ket{\rho_0}. \nonumber
 \end{align} 
    For an observable $G_n = \bra{g}\ket{\rho_n}$, we therefore obtain
    \begin{equation}
   G_n =  \sum_{k = 0}^n  \bra{g} (Q \Lambda)^k \ket{\xi_{n-k}} +\bra{g} (Q \Lambda)^n Q \ket{\rho_0}.
   \label{prediction}
   \end{equation}
   In the Heisenberg picture, $G_n = \bra{g_n}\ket{\rho_0}$, and from \eqref{prediction} it follows that 
   \begin{align}
   \label{adjoint-predict}
   \bra{g_n} & = \sum_{k = 0}^n  \bra{g} (Q \Lambda)^k P \Lambda^{n-k} + \bra{g} Q (\Lambda Q)^n \\
   & = \bra{g} P \Lambda^n + \sum_{k=1}^n \bra{g} Q (\Lambda Q)^{k-1} \Lambda P \Lambda^{n-k}  + \bra{g} Q (\Lambda Q)^n \nonumber \\
   & = \bra{g} P \Lambda^n + \sum_{k=1}^n \bra{F_{k-1}} \Lambda P \Lambda^{n-k} + \bra{F_n}, \nonumber 
   \end{align}
   where we have  defined $\bra{F_k} = \bra{g}Q (\Lambda Q)^k$. Equation (\ref{adjoint-predict}) is 
identical to Equation~(6) in Ref.~\cite{Darve_Computing_2009}. It follows that $\bra{F_k}\ket{\xi_j} = 0$ for all $j$ and $k$ since $QP = 0$. For this reason, $\bra{F_k}$ is usually treated as `noise', although, in principle, one can characterize $\bra{F_k}$ through solutions of the  the {\em orthogonal dynamics} \cite{Givon_Existence_2005,Darve_Computing_2009} (See also Appendix~\ref{sec:orthogonal})
 \begin{equation}
 \label{orthogonal}
    \bra{F_{n+1}} = \bra{F_n} \Lambda Q, \qquad \bra{F_0} = \bra{g} Q.
    \end{equation}

 Equation~\eqref{adjoint-predict} is the discrete time Mori-Zwanzig decomposition and~\eqref{master-eqn} below is the adjoint,  which evolves  the states  instead of the observables. These equations are identities and are often taken as starting points for building methods to estimate quantities that are not directly observed, in terms of quantities $\xi_{n}, \xi_{n-1}, \ldots, \xi_0$ that have been observed by time $n$.
    
    A problem of significant interest is {\em prediction}, i.e. estimating $\xi_{n}$ using the information available at time $n-1$, which are the quantities $\xi_{n-1}, \xi_{n-2}, \ldots, \xi_0$. Using.~\eqref{mz_discrete} with $n \to n-1$ and $\ket{\xi_{n}} = P \Lambda \ket{\rho_{n-1}}$, we obtain
    \begin{equation}
    \ket{\xi_{n}} = \underbrace{P \Lambda   \ket{\xi_{n-1}}}_{\text{Markovian}} + \underbrace{\sum_{k=2}^n P \Lambda (Q \Lambda)^{k-1}  \ket{\xi_{n-k}}}_{\text{memory}} + \underbrace{P \Lambda (Q \Lambda)^{n-1} Q \ket{\rho_0}}_{\text{noise}},
    \label{master-eqn}
    \end{equation}
  where  the right hand side is decomposed into the Markovian term, the ``optimal" estimate of $\ket{\xi_{n}}$ given the current state $\ket{\xi_{n-1}}$, the memory term that encodes the dependence on the past observations $\ket{\xi_{n-2}}, \ket{\xi_{n-3}}, \ldots, \ket{\xi_0}$, and the noise, which is orthogonal to $\ket{\xi_j}$ and depends on the microscopic details of the initial condition, i.e it depends on $\ket{\rho_0}$ and not just on $\ket{\xi_0}$. Of course, an important caveat here is that the interpretation of the decomposition as Markovian, memory and noise terms relies on the origins of this procedure in the near-equilibrium statistical mechanics context, and {\em it is by no means clear that this interpretation is valid for the evaporation process \eqref{eq:continuum}.}  
  
  Nonetheless, the equation is a formally exact decomposition of a PDE with stochastic initial conditions into a part that only depends on a subset of the degrees of freedom, {\em the resolved variables}, along with an exact expression for the  remainder. In what follows, we will lump the Markovian and the memory terms into a single quantity, so the distinction between the resulting two terms is whether or not they only depend on the observed quantities $\ket{\xi_j}$, or the entire (microscopic) initial condition $\ket{\rho_0}$.

\subsection{The Memory Kernel for the Weathering of Oil}

We now return to the  evaporation model. We assume there is  a single observed quantity, $M_n = \int \rho_n dw$. We take $\mathcal{H} = L^2([0,1])$, the space of square integrable functions on $[0,1]$. Defining $\ket{1}$ to denote the constant function $g(x) = 1$, we have 
$$
M_n = \bra{1}\ket{\rho_n}, \qquad \bra{1}\ket{1} = 1.
$$
The orthogonal projection on to the one-dimensional space spanned by the constant functions is given by $P = \ket{1}\bra{1}$. Consequently, $\ket{\xi_n} = P \ket{\rho_n} = \ket{1} \bra{1}\ket{\rho_n} = M_n \ket{1}$ and $M_n = \bra{1}\ket{\xi_n}$. Using this in~\eqref{master-eqn} and projecting on to constants gives
\begin{equation}
\label{mz2}
M_{n} = \sum_{k=1}^n h_k M_{n-k} + \beta_n,
\end{equation}
where $h_k = \bra{1} (\Lambda Q)^{k-1} \Lambda \ket{1}$ and $\beta_n = \bra{1} (\Lambda Q)^{n} \ket{\rho_0}$ is the `noise' that depends explicitly on the microscopic initial condition $\ket{\rho_0}$.  We can also obtain the same equation from the (usual) Mori-Zwanzig  decomposition in~\eqref{adjoint-predict} by taking $\bra{g} = \bra{1}\Lambda$.

Equation~\eqref{mz2} {\em is exact}. In particular, it holds for $\ket{\rho_0} = \ket{1}$, in which case the noise vanishes, $\beta_n \equiv 0$ for all $n$. If $\ket{\rho_0} = c \ket{1}$, for some constant $c$, $M_0 = \bra{1}c\ket{1} = c$ and we can explicitly solve \eqref{eq:continuum} to obtain 
\begin{equation}
M_n = M_0\int_0^1 e^{-w n \tau} dw = \begin{cases} M_0\frac{1 - e^{-n \tau}}{n \tau} & n \geq 1, \\ M_0 & n = 0. \end{cases}
\label{open-loop}
\end{equation}
Consequently, the memory kernel $h_k$ is determined by 
\begin{equation}
 \frac{1 - e^{-n \tau}}{n \tau} =  \sum_{k=1}^n h_k  \frac{1 - e^{-(n-k) \tau}}{(n-k) \tau} \qquad \mbox{ for all } n \geq 1.
 \label{convolve}
 \end{equation}
 We can solve for $h_k$ using the $\mathcal{Z}$-transform (equivalently the generating function). Let $\hat{M}(z) = \sum_{n = 0}^\infty M_n z^{-n}$ and $\hat{H}(z) = \sum_{n = 0}^\infty h_n z^{-n}$. The sum defining $\hat{M}(z)$ converges for $z$ outside the unit disk since $M_n$ is clearly a decreasing sequence. We can compute the sum explicitly to obtain
 $$
 \hat{M}(z) = M_0 \sum_{n=0}^\infty \int_0^1 e^{-w n \tau} z^{-n} dw = M_0\int_0^1 \frac{z e^{w \tau}}{ze^{w \tau}-1} dw = \frac{M_0}{\tau} \log\left[\frac{ze^\tau-1}{z-1}\right].
 $$
 Multiplying~\eqref{convolve} by $z^{-n}$ and summing on $n \geq 1$ gives
 \begin{equation}
 \hat{M}(z) -M_0 = \hat{M}(z) \hat{H}(z),
 \label{functional}
 \end{equation}
 and rearranging yields
 \begin{equation}
 H(z) =  \left[1 - \frac{M_0}{\hat{M}(z)}\right] =  \left[1 - \frac{\tau}{\log(e^\tau z -1) - \log(z-1)}\right].
 \label{Hz}
 \end{equation}
$H$ is analytic outside the unit circle and has a branch point singularity at $z = 1$. Expanding about $z = \infty$ gives 
 $$
 H(z) = z^{-1} \frac{1-e^{-\tau}}{\tau} + z^{-2} \frac{(1-e^{-\tau})((\tau-2) +(\tau+2)e^{-\tau})}{2\tau^2} + \cdots,
 $$
 so that the coefficients $h_k$ can be explicitly computed. Our interest is in the long time behavior of $h_k$, which can be deduced from  the $z \to 1$ behavior of $\hat{H}(z)$. $\hat{H}(z)$ has a logarithmic branch point at $z = 1$. In particular, this implies that the series for $\hat{H}(z)$ does not converge for any $z$ with $|z| < 1$, so that the sequence $h_k$ decays slower than the exponential $e^{-\epsilon k}$ for any $\epsilon > 0$. The transfer operator methods in Flajolet and Odlyzko \cite{Flajolet_Singularity_1990} (Theorem 3A and comments on pp. 231--232) imply, in fact, that 
 \begin{equation}
 h_k \sim \frac{1}{k \log^2(k)} \qquad \mbox{ as } k \to \infty,
 \label{long-tail}
 \end{equation}
 so that $h_k$ decays algebraically. Although $\sum h_k$ converges to $\hat{H}(1) = 1$ (by \eqref{Hz}), the partial sums go to 1 extremely slowly, $\left|1-\sum_{k=1}^N h(k)\right| \sim \log(N)^{-1}$.
 
 The memory kernel $h_k$ thus has a fat tail.  The algebraic decay of $h_k$ is a reflection of the extremely slow relaxation in $\rho(w,t) = \rho_0(w) e^{-wt}$ for species with $w$ close to zero. In general, the initial condition $\rho_0(w)$ has an effect for times of order $1/w$, so the initial condition is not ``forgotten" for long times, leading to the fat tails and slow decay of correlations. 
              
              \section{Dimension Reduction, Stochastic Modeling and Filtering} \label{sec:parametrization}

We are interested in {\em model reduction}, i.e. in developing  low dimensional  (approximate) models for predicting the behavior of high dimensional complex systems, e.g. the linear evaporation process~\eqref{eq:continuum}. Before we describe our work on this problem, we first review some basic terminology from signal processing, and then present a roadmap to guide the reader through our various approaches to the problem of prediction/model reduction for~\eqref{eq:continuum}.

\subsection{Filtering, estimation and prediction}

In our context, the general prediction/estimation problem along with data assimilation is the following: The sequence $\ket{\rho_k}$ describes the ``state" of the system~\eqref{eq:continuum} sampled at discrete times $t = k \tau$. We are given a sequence of noisy measurements
      $
     \displaystyle{ \tilde{M}_{k} = \braket{1}{\rho_k} + \sigma \gamma_k}$ where the  $\gamma_k$ are uncorrelated normal variates. What is the ``best" prediction for $M_n = \braket{1}{\rho_n}$ in terms of the measurements $\tilde{M}_k$ for $k < n$? Abstractly, the optimal estimate is given by a conditional expectation 
     $$
     \bar{M}_n = \mathbb{E}[M_n \, | \, \tilde{M}_{n-1}, \tilde{M}_{n-2},\ldots,\tilde{M}_{1},\tilde{M}_{0}].
     $$
     We seek a concrete representation for the optimal estimator, i.e. a (sequence of) explicit  functions $F_n$ such that 
     $$
      \mathbb{E}[M_n \, | \, \tilde{M}_{n-1}, \tilde{M}_{n-2},\ldots,\tilde{M}_{1},\tilde{M}_{0}] \approx F_n(\tilde{M}_{n-1}, \tilde{M}_{n-2},\ldots,\tilde{M}_j,\ldots).
      $$
     with $\tilde{M}_j = 0$ for $j \leq 0$. We will call such functions $F_n$  {\em filters} or {\em predictors}. We can classify filters by the following properties:
     \begin{enumerate}
     \item The filter is {\em autonomous} or {\em shift-invariant} if $F_n \equiv F$  independent of $n$.
     \item If $F_n$ only depends on $\tilde{M}_{n-1}, \tilde{M}_{n-2},\ldots,\tilde{M}_{n-L}$ for some finite $L$, then it is a {\em finite impulse response (FIR)} filter with $L$ {\em taps}. Otherwise, the filter is an {\em finite impulse response (IIR)} filter that uses information from the entire time history of the time series $\tilde{M}_k$.  
     \item A filter $F_n$ is {\em linear} if it is given by a linear function of its arguments.
     \item A filter $F_n$ is {\em genie-aided} if it has access to more information than is available in $\tilde{M}_{n-1}, \tilde{M}_{n-2},\ldots$. Such filters cannot be built in practice. Nonetheless, as with the Maxwell demon, this fictional construct is useful because it allows us to bound the best-case behavior of constructible filters.
     \item A filter $F_n$ is {\em empirical} or {\em data-driven} if it is obtained through regression on many realizations of the underlying random process $\tilde{M}_k$.
     \end{enumerate}
     
\begin{table}[htbp]
   \centering
   \begin{tabular}{@{} lcccr @{}} 
      \toprule
Filter & Definition  & Linear & Data driven &  Other features  \\
      \midrule
      \\
      \multicolumn{3}{l}{Memory-kernel/Transfer function based methods} \\
      \cmidrule(r){1-3} 
      \\
      MZ filter & Eq.~\eqref{eq:MZestimator} & Yes  & No & Needs all history \\
      Truncated MZ (FIR) & Eq.~\eqref{truncated_FIR} & Yes & No &  \\
      MZ-Pade & Eq.~\eqref{poly-estimator} & Yes  & No &  \\
      Harmonic Filter & Eq.~\eqref{harm-estimator} & No & No & \\
      \midrule
      \\
      \multicolumn{2}{l}{Statistical regression based methods} \\
      \cmidrule(r){1-2} 
      \\
      Linear Oracle & Sec.~\ref{sec:empirical}  & Yes  & Yes & Genie-aided \\
      Empirical Linear & Eq.~\eqref{eq:bavg} & Yes  & Yes & Averaged over runs \\
      Empirical Harmonic & Eq.~\eqref{eq:havg} & No  & Yes &  Averaged over runs\\
      \midrule
     \\
      \multicolumn{3}{l}{Methods that exploit slow decay of correlations} \\
      \cmidrule(r){1-3} 
      \\
      Asymptotic filter & Eq.~\eqref{asymp-filter} & Yes &  No & Non-autonomous \\
      Universal filter & Eq.~\eqref{universal} & Yes &  No & Unstable \\
      Extended asymptotic & Eq.~\eqref{eaf} & No & No & Hidden variables\\ 
      \bottomrule
   \end{tabular}
   \caption{A summary of the various filters we will consider in this work, along with a description of their features. All but the Asymptotic filter are shift-invariant, and all but the MZ filter have finitely many taps.}
   \label{tab:all_filters}
\end{table}

Table~\ref{tab:all_filters} describes our various approaches to building filters for the linear evaporation process~\eqref{eq:continuum}. At the gross level, there are three distinct approaches. The first approach is based on the memory kernel \eqref{Hz}, or equivalently, the single realization corresponding to $\ket{\rho_0} = \ket{1}$ given by \eqref{open-loop}. These filters are described in sections~\ref{sec:linear}~and~\ref{sec:nonlinear}. The second approach, discussed in section~\ref{sec:empirical}, is empirical and relies of estimating coefficients in filter functions using statistical regression on independent realizations of the random process~\eqref{eq:continuum}. The final approach is non-empirical, and exploits the slow relaxation inherent in the process $M_k$. In this case, the slow decay of the memory kernel~\eqref{long-tail} is beneficial, rather than detrimental, contrary to intuition. This approach is discussed in section~\ref{sec:analysis}.

We also note that solving the filtering/prediction problem is very closely related to obtaining reduced models for the high-dimensional system~\eqref{eq:continuum}. If $F_n$ is a (close to) optimal filter, then the process 
$$
\widehat{M}_n =  F_n(\widehat{M}_{n-1}, \widehat{M}_{n-2},\ldots,\widehat{M}_j,\ldots) + \theta_n,
$$
 where the quantities $\theta_n$ are stochastic with the appropriate statistics, is a good surrogate for the high dimensional process that generates $M_n$. This reduction is particularly efficient if the filter $F_n$ is shift-invariant and has finitely many taps. Indeed, this is the framework in which the Mori-Zwanzig projection operator formalism is used to build reduced models for various high-dimensional systems \cite{Chorin_Discrete_2015,Lu_Data_2017,Harlim_Regression_2013}.
 
 A natural question is: ``{\em Why consider multiple approaches?}" We do this because  we have good analytical understanding of the `high-dimensional' dynamics of~\eqref{eq:continuum}, so we have a good theoretical basis for assessing the performance of many of the popular approaches to stochastic modeling/dimension reduction. We are able to evaluate the relative merits of the various assumptions/approximations that are inherent in the different approaches. Finally, we are able to develop an analytic framework that gives new approaches to model reduction for high dimensional systems with long term memory, and one that is applicable to practical problems.
 
In order to assess the performance of our various filters, as well as to generate the empirical filters by regression, we need realizations of the evaporation process~\eqref{eq:continuum} with random initial data $\ket{\rho_0}$ drawn from an appropriate distribution. We numerically generate such realizations as follows: \\

\noindent {\bf Algorithm I}: Generating synthetic data \\
\label{alg1}
\begin{enumerate}
\item We discretize the interval $[0,1]$ into $I$ equal intervals of size $\Delta = 1/I$. For our simulations we take $I = 1000$.
\item We assign the initial mass distribution by picking $I$ independent random variables $u_i, i = 1,2,\ldots,I$ uniformly distributed random variables on $[0,1]$ and then normalize to set 
$$
\rho_0(i) =  \frac{u_i}{\Delta \sum_{i=1}^I u_i}.
$$
By symmetry, the marginal distributions of the quantities $\rho_0(i)$ are identical, but they do depend on $I$, the ``total number of species" . They are however not independent. By construction $\sum_{i} \rho_0(i) = I$.
To the interval $[(i-1)\Delta,i \Delta]$, indexed by $i$, we associate the decay rate 
$$
w(i) = \left(i-\frac{1}{2}\right)\Delta, \qquad i = 1,2,\ldots I,
$$
corresponding to the middle of the interval. 
\item We pick $\tau = \log(3/2)$ so that $\rho_n(i) = \rho_0(i) \left(\frac{2}{3}\right)^{n w(i)}$.
\item We compute $M_n = \sum \rho_n(i) \Delta$ for $1 \leq n \leq N$, where we choose $N$ such that the assumption that we are discretizing a continuum density using $I$ intervals is still valid. This requires that there are at least $\sim 10$ intervals for which the density $\rho_n(i)$ has not decayed down to zero. This gives the rule of thumb $\log(3/2) w(10) N \sim 1$ so that $N \sim \frac{I}{10 \log(3/2)} \sim \frac{I}{4}$. We can thus safely take $N = 200$.  \qed
\end{enumerate}

The numerical procedure is very close in spirit to the original discrete model~\eqref{eq:species} with $I$ distinct species. The one difference is that, since we are rediscretizing a continuum limit, we can pick the decay rates $w(i)$ on the basis of our discretization, and not through any relation with the ``true" decay rates of the components of oil. This also ties in with the idea that the individual decay rates in the mixture cannot be identified, and our methods have to be robust to possible changes in the underlying ``bare" decay rates.

Through this procedure we obtain many random realizations of (a discretization of) the system in~\eqref{eq:continuum} with $M_0 = 1$. We use the computed values of $M_k$ as the ``measurements" $\tilde{M}_k$ in estimating $M_n$ from the measurements for $k < n$.  In particular, we will assume there is no measurement noise.

 \subsection{The Mori Projection and Linear Autonomous Estimators} \label{sec:linear}
  
  Equation~\eqref{mz2} is {\em exact} (see also~\eqref{mz3} in Appendix~\ref{sec:multi}) 
  and gives a stochastic reduced model of the system~\eqref{eq:continuum} on replacing the quantities $\beta_n$ (determined by the microscopic initial conditon $\ket{\rho_0}$) with a stochastic process $\theta_n$, typically a Gaussian process, that has the same ``statistics", i.e.  we match the means and the covariances
  $$
  \mathbb{E}[\beta_n] = \mathbb{E}[\theta_n] = 0, \qquad  \mathbb{E}[\beta_n^T \beta_m] = \mathbb{E}[\theta_n^T \theta_m], \qquad \mbox{ for all } m,n \geq 0,
  $$
where the expectations $\mathbb{E}$  for $\beta$ are over a natural measure for the initial conditions and the expectations for $\theta$ are over the measure underlying the stochastic process $\theta_n$. Equation~\eqref{mz2} thus gives the stochastic model                       
      $$\displaystyle{
      M_{n} = \sum_{k = 1}^n h_k M_{n-k} + \theta_n.}$$
 Consequently, we also have the associated prediction/filtering algorithm
  \begin{equation}
\bar{M}_{n} = \sum_{k= 1}^n h_k \tilde{M}_{n-k},
\label{eq:MZestimator}
    \end{equation}
which we will call the MZ filter. 
The MZ filter is shift-invariant (autonomous), but nonetheless evaluating the sum in \eqref{eq:MZestimator} requires us to keep track of the entire history of $\tilde{M}_n$, As we argued above, since $h_k$ has a fat tail, one cannot simply truncate the sum at a fixed $L$ and expect to get good results. 

\begin{figure}[tbhp] 
         \centering
         \includegraphics[width=0.95 \textwidth]{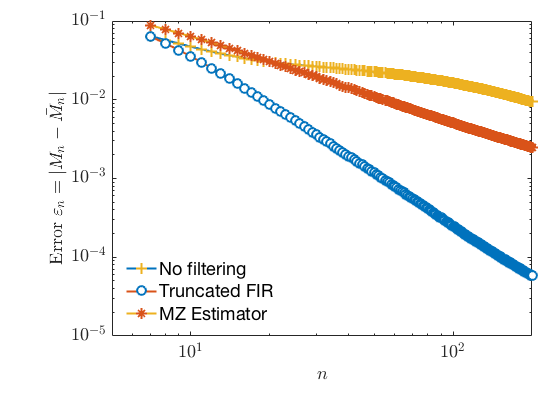} 
         \caption{The inferred error for the three estimators used on one random realization. The estimator with no data assimilation (yellow) is given by~\eqref{open-loop}, the renormalized FIR filter  (blue) is given by~\eqref{truncated_FIR} and the Mori-Zwanzig estimator (red) that uses the entire history is given by~\eqref{eq:MZestimator}.}
         \label{fig:simple}
      \end{figure}

Figure~\ref{fig:simple} compares the performance of three potential estimators. The first estimator does not use any data assimilation, so the predicted sequence $\bar{M}_{n}$ is given by \eqref{open-loop}. The second estimator truncates the sum in the Mori-Zwanzig decomposition at $L = 6$ where the value 6 has no particular significance and is chosen purely for the purposes of illustrating the effects of truncating the sum. Naive truncation gives the estimator  $
\displaystyle{ \bar{M}_{n} \approx \sum_{k= 1}^{L} h_k \tilde{M}_{n-k}
      }$, which is biased at $O(n^{-1})$, because the quantities $h_{k} M_{n-k}$ have positive means and their sum over $n-L \leq k < n$ is $O(n^{-1})$. We can attempt to eliminate this bias by an {\em ad hoc} ``renormalization" of the the weights 
  \begin{equation}
  \label{truncated_FIR}
      h'_k = \frac{h_k}{\sum_{k=1}^{L} h_k}, \quad \bar{M}_{n} = \sum_{k= 1}^{L} h'_k \tilde{M}_{n-k}
 \end{equation}
      so that $\displaystyle{\sum_{k=1}^{L} h'_k = \sum_{k=1}^{\infty} h(k) = 1}$, and the estimator has a bias $O(n^{-2})$ and is thus `better' in the limit $n \to \infty$. We will call this the {\em truncated FIR} (Finite impulse response) filter in contrast to the third estimator, \eqref{eq:MZestimator} which is an {\em IIR} (Infinite impulse response) filter that is obtained from the Mori-Zwanzig decomposition and incorporates the entire history of $M_n$.  
      
\begin{figure}[tbhp] 
         \centering
         \includegraphics[width=0.95 \textwidth]{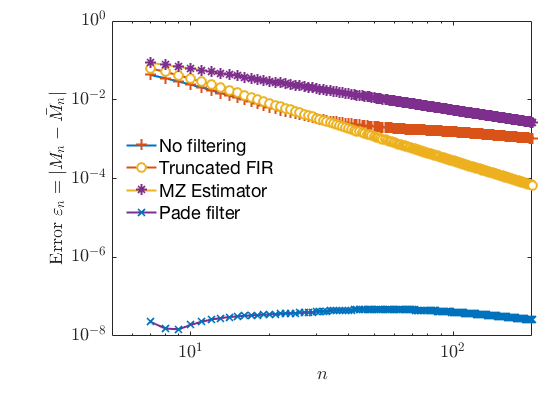} 
         \caption{The inferred error averaged over 100 realizations. These curves are very stable and do not vary discernibly between sets of 100 independent realizations. Three of the estimators are the same as in \protect{Fig.~\ref{fig:simple}}. We also compare the FIR filter generated by the [6,6] Pad\'{e} approximant of the Mori-Zwanzig transfer function $H(z)$. The one-step prediction error for the Pad\'{e} filter  is substantially smaller than the errors in the other estimators. }
         \label{fig:error1}
      \end{figure}

      For each estimator, we define the  ``inferred noise" or the one-step prediction error $\varepsilon_n$  as the difference $|\bar{M}_{n}-M_n|$ between the estimate $\bar{M}_n$ using information available at time $n-1$ and the (random) value $M_n$ (``the truth") for  a realization. We display these differences for a single `typical' realization in fig.~\ref{fig:simple}. Figure~\ref{fig:error1} show the averaged error over many realizations
       
Very surprisingly, the MZ estimator using the entire history does worse than the truncated MZ estimator with renormalization and, on average, {\em also worse than the estimator without any data assimilation}.

 An alternative approach to truncating the sum in the Mori-Zwanzig estimator is to work in the $\mathcal{Z}$-transform domain and approximate $\hat{H}(z)$ in~\eqref{functional} by 
a rational function in $z^{-1}$ \cite{dsp}, i.e. $\hat{H}(z) \approx p(z^{-1})/q(z^{-1})$ where $p$ and $q$ are polynomials  of degrees less than or equal to $L$, and we normalize by requiring that $q(0) = 1$. Since $h_0 = 0$, it follows that $p(0) = 0$. The $\mathcal{Z}$-transform of the sequence $\bar{M}_n$ of estimates given by 
$$
\sum_{n=0}^\infty \bar{M}_n z^{-n} \equiv \widehat{\bar{M}}(z) \approx \frac{M_0}{1-\frac{p(z^{-1})}{q(z^{-1})}} = M_0 \frac{q(z^{-1})}{q(z^{-1})-p(z^{-1})}.
$$
If we define $b(z^{-1}) = q(z^{-1})-p(z^{-1})$, $b$ is also a polynomial with the normalization $b(0) = 1$. Thus, we get an auto-regressive AR($L$) model \cite{box-jenkins,regression} $b(z^{-1})  \widehat{\bar{M}}(z) = M_0 q(z^{-1})$. Writing $b(z^{-1}) = 1 + b_1 z^{-1} + b_2 z^{-2} + \ldots + b_j z^{-j}$ and $q(z^{-1}) = 1 + q_1 z^{-1} + q_2 z^{-2} + \ldots + q_j z^{-j}$, we have the estimator
$$
\bar{M}_{n} = M_0 q_{n} - \sum_{k=1}^{n} b_{k} \bar{M}_{n-k} ,
$$
where we have used the convention $b_k = 0$  (resp. $q_k = 0$) for indices $k$ greater than the degrees of the respective polynomials. The sum on the right hand side therefore has no more than $L$ non-zero terms. Of course, if the estimator is `good', then $\bar{M}_n$ the estimate for $M_n$ using information available prior to time $n$ is close to the true value $M_n$. In fact, one might argue that the Mori-Zwanzig decomposition~\eqref{mz2} is exact for the ``true" sequence $M_n$ and thus one would do better (or certainly not much worse) by replacing the estimates $ \bar{M}_{n-k} $ by their measured values $M_{n-k}$  since we are assuming there is no measurement error. This gives the estimator
\begin{equation}
\bar{M}_{n} = M_0 q_{n} - \sum_{k=1}^{n} b_{k} M_{n-k}.
\label{poly-estimator}
\end{equation}
One can view this as an alternate renormalization of the weights $h_k$ in a truncated MZ estimator, one that is perhaps better justified and less ad hoc than the choice $h'_k = h_k (\sum_{k=0}^{L} h_k)^{-1}$ from above. 
In Fig~\ref{fig:error1}, we present the results using the [6,6] Pad\'e approximant of $H(z)$ about $z = \infty$ to obtain the 6th order polynomials $p$ and $q$, which then give a 6 tap filter~\eqref{poly-estimator} for predicting $M_n$ from $M_{n-1},M_{n-2},\ldots,M_{n-6}$. Very surprisingly, the Pad\'e filter, which is ostensibly designed to approximate the MZ estimator through a filter with finitely many delays, {\em performs significantly better} than the MZ estimator \eqref{eq:MZestimator}.

              \subsection{Nonlinear Filters} \label{sec:nonlinear}
              
              Since the evaporation process~\eqref{eq:continuum} is linear, the equations do not have a natural scale for the mass $M$. In particular, the total amounts of oil is an extensive quantity whose (scale-free) logarithmic derivative $\partial_t M(t)/M(t)$ should only depend on the relative fractions of the different species and not their total amounts. It follows that the dynamics, and consequently also the estimators, should be homogeneous of degree 1, i.e if the estimator of $M_n$ is  given by $\bar{M}_{n} = F_n(M_{n-1},M_{n-2}, \ldots,M_1,M_0)$, then $F_n$ must satisfy
              $
              F_n(\lambda M_{n-1}, \ldots,\lambda M_1,\lambda M_0) = \lambda F_n(M_{n-1}, \ldots,M_1,M_0)$ for all  $\lambda > 0.
              $
     It is easy to verify that all the linear estimators in section~\ref{sec:linear} have this property.
     
     Note also that the Mori-Zwanzig decomposition~\eqref{mz2} is exact, and further, the noise in the equation is exactly zero for choices of the initial condition $\ket{\rho_0}$ in the range of the projection $P = \ketbra{1}{1}$. Consequently, the estimators in the previous section were designed to (exactly or approximately) recover the sequence $M_n$ for initial conditions in the range of $P$. We can also seek nonlinear estimators with the same property. 
     
     The motivation for considering nonlinear estimators that the process $M_{n}$ behaves (roughly) like $1/n$, so no finite-lag ARMA filter can generate such a process. Indeed, the impulse response of such a filter is determined by its poles \cite{dsp} and consists of sums of sequences of the form  $n^l z_j^{-n}$ where $l < $ the order of the pole $z_j$ of the filter transfer function. On the other hand, for the sequence in \eqref{open-loop}, $M_{n}^{-1} \sim n$ with exponentially small corrections. The sequence $M_n^{-1} = n$ can indeed be generated by an ARMA filter \cite{box-jenkins}, in particular, by a 2nd order pole at $z_j = 1$. 
     
      This naturally leads us to filter $M_n^{-1}$ by seeking weights $\nu_k$ such that 
     $$
     \frac{1}{M_{n}} = \sum_{k = 1}^n \frac{\nu_k}{M_{n-k}}
     $$
     {\em exactly} for the sequence~\eqref{open-loop}, and then use this as a starting point for truncation as in the previous section~\ref{sec:linear}. We have, for $\ket{\rho_0} = \ket{1}$,
  \begin{equation}
     \frac{1}{M_{n}} = \begin{cases} \frac{n \tau}{M_0(1 - e^{-n \tau})} = \frac{n\tau}{M_0} + \frac{n \tau e^{- n \tau}}{M_0} + \frac{n \tau e^{- 2 n \tau}}{M_0} + \cdots & n \geq 1, \\ \frac{1}{M_0} & n = 0. \end{cases}
\label{harmonic-open-loop}
\end{equation} 
The $\mathcal{Z}$-transform of the sequence $M_n^{-1}$ is given by 
$$
T(z) = \sum_{n = 0}^\infty z^{-n} \frac{1}{M_n} = \frac{1}{M_0}\left[1 + \frac{z^{-1} \tau}{(1-z^{-1})^2} + \sum_{j = 1}^\infty \frac{z^{-1} \tau e^{-j \tau}}{(1-e^{-j \tau} z^{-1})^2}\right].
$$
Note that, in contrast to the $\mathcal{Z}$-transform for $M_n$, which has logarithmic branch points at $z=1$ and $z=e^{-\tau}$, the $Z$-transform for $M_{n}^{-1}$ is has poles of order 2 at $z = e^{-j \tau}, j = 0,1,2,\ldots$ and an essential singularity at $z = 0$. 

As before, we can compute the sequence $\nu_k$ through its $\mathcal{Z}$-transform by
$$
S(z) = \sum_{k=0}^\infty \nu_k z^{-k} = z\left[1 - M_0^{-1}T(z)^{-1}\right] .
$$
This form is not directly useful, since we do not know the zeros of $T(z)$, which correspond to the poles of $S(z)$, which in turn determine the asymptotic behavior of $\nu_n$. 

\begin{figure}[htbp] 
         \centering
        \includegraphics[width=0.95 \textwidth]{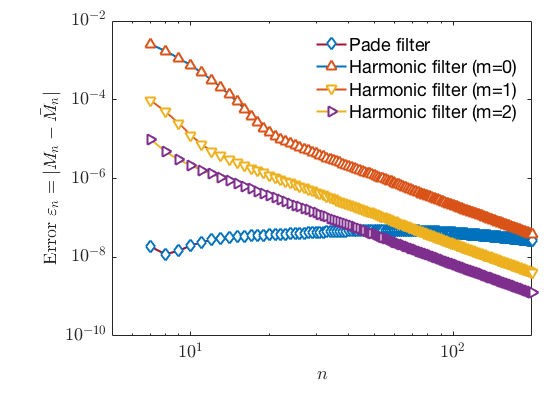} 
         \caption{The one step error for  the harmonic filters with $m=0,1$ and $2$ corresponding to predictions using $L=2,4$ and $6$ respectively. We also show the inferred error of the Pad\'{e} filter which was the `best' among the various linear filters that we considered in~\ref{sec:linear}}
         \label{fig:harm}
      \end{figure}

We will instead use an alternative approach that directly determines rational approximations to $T(z)$ which can then be used to generate finite lag estimators for $M_n^{-1}$. Since the series expansion for $\frac{1}{M_n}$ in \eqref{harmonic-open-loop} converges exponentially, we can truncate  the sum at order $e^{-m n \tau}$ to obtain
$$
M_0T(z) \approx  1 + \frac{z^{-1} \tau}{(1-z^{-1})^2} + \frac{z^{-1} \tau e^{-\tau}}{(1-e^{-\tau} z^{-1})^2}
+ \cdots + \frac{z^{-1} \tau e^{- m\tau}}{(1-e^{-m \tau} z^{-1})^2},
$$
 a rational function approximation, where the error in this approximation is uniformly bounded on the unit circle. Given a rational approximation to $T(z) = M_0^{-1} q(z^{-1})/b(z^{-1})$, we can design a linear predictor for $M^{-1}_{n}$ as above (cf. \eqref{poly-estimator}). We illustrate this method with explicit calculations for  $m = 1$:
 \begin{align}
 \label{4tap}
T(z) & \approx \frac{1}{M_0} \left[1 +    \frac{z^{-1} \tau}{(1-z^{-1})^2} + \frac{z^{-1} \tau e^{-\tau}}{(1-e^{-\tau} z^{-1})^2} \right]   \nonumber \\
 & = \frac{1}{M_0} \frac{( (1-z^{-1})^2 (1-e^{-\tau} z^{-1})^2 + z^{-1} \tau \left[(1-z^{-1})^2 + (1-e^{-\tau} z^{-1})^2\right]}{(1-z^{-1})^2 (1-e^{-\tau} z^{-1})^2} \nonumber \\
 & \equiv \frac{1}{M_0}\frac{q(z^{-1})}{b(z^{-1})},
\end{align}
where the last line defines the polynomials $b$ and $q$ through the expressions on the middle line. The polynomials $b$ and $q$ are normalized, $b(0) = q(0) = 1$, and have degree $2m+2$ (in general)  corresponding to the $m+1$ quadratic factors from the poles of order $2$ in the rational approximation of $T(z)$. By the same arguments as in sec~\ref{sec:linear}, we get the following estimator for $M_{n}^{-1}$:
\begin{equation}
\bar{M}^{-1}_{n} = M^{-1}_0 q_{n} - \sum_{k=1}^{n} b_{k} M^{-1}_{n-k},
\label{harm-estimator}
\end{equation}
with the convention that $b_j =q_j =0$ for $j > 2m+2$  so that the sum on the right hand side has at most $2m+2$ non-zero terms, i.e 4 terms in the case in \eqref{4tap} with $m = 1$. We will refer to the estimator in \eqref{harm-estimator} as a {\em harmonic filter}, because for the case $m =0$, the filter reduces to 
$$
\bar{M}_{n}^{-1} = 2 M_{n-1}^{-1} - M_{n-2}^{-1} \qquad \mbox{ for }n \geq 3,
$$
i.e. the middle value $M_{n-1}$ is the harmonic mean of the extreme values $M_{n-2}$ and $M_{n}$. Fig.~\ref{fig:harm} shows the comparison between the performance ofthe pade-truncated MZ estimator \eqref{poly-estimator} with 6 taps, and the harmonic estimator  \eqref{harm-estimator} with $m=0,1$ and $2$. The Pad\'e truncated estimator performs better for small $n$, while the Harmonic predictors have comparable or superior performance for $n \gtrsim 30$. 

\subsection{Empirical Filters: Data-Driven Stochastic Parameterization} \label{sec:empirical}
  
A powerful approach to stochastic parameterization  is through data-driven model reduction \cite{Chorin_Problem_2007,Chekroun_Predicting_2011,Harlim_Regression_2013,Lu_Data_2017}. In this approach,  the coefficients in a parametric or semi-parametric model are determined by comparison with data. For this procedure, it is crucial that one begins with an appropriate form for the reduced model. Indeed a model with too many parameters can overfit the training data, i.e. the coefficients can become sensitive to the random noise in the data, which will then lead to poor predictions from the reduced model.  We can view the finite lag linear and harmonic estimators in the previous two sections (see eqs.~\eqref{poly-estimator}~and~\eqref{harm-estimator}) as useful {\em ansatzes} for building data driven `empirical' estimators. In this approach, the coefficients of the polynomials $q$ and $b$ are not inferred from the particular sequence corresponding to the initial condition $\ket{\rho_0} = M_0 \ket{1}$, but rather, are estimated from many (random) realizations of the time series for $M_n$ with the initial condition $\ket{\rho_0}$ sampled from an appropriate distribution.     

The appropriate state space model is $M_{n} = \sum_{k=1}^n h_k M_{n-k} + \beta_n$ where $\beta_n = \bra{1} (\Lambda Q)^{n} \ket{\rho_0}$ is now a {\em non-stationary} random process (See sec.~\ref{sec:analysis}). Nonetheless, as a crude approximation, we can assume the AR($L$) model 
$$
\bar{M}_{n} = M_0 q_{n} +\sum_{k=1}^{\min(n,L)}   h'_k M_{n-k} + \sigma(\{M_{n-1},M_{n-2},\ldots,M_0\}) \theta_n,
$$ where $q_k$, $h'_k$ are fixed (in $n$) renormalized weights that are zero for $n > L$, and $\theta_n$ are i.i.d. normal variates, giving a stochastic parameterization \cite{Arnold_Stochastic_2013} of $M_n$. The non-stationarity of the noise process is modeled through the variance parameter $\sigma$ that could, in principle, depend on the state as encoded by the entire history $\{M_n,M_{n-1},\ldots,M_0\}$. This is a common idea in regression analysis, called {\em variance inflation} \cite{regression}. It is an uncontrolled approximation because we are insisting that the covariance matrix for the fluctuations be diagonal. 

We estimate $\sigma$ by (i) using the homogeneity of the underlying process \eqref{eq:continuum}, and (ii) assuming that, for $n > L$,  $\sigma_n$ only depends on the `recent' past of $M_n$ so it is only a function of the quantities $M_{n},M_{n-1},\ldots,M_{n-L}$ that appear in the sum $\sum_{k=1}^{L}  h'_k M_{n-k}$.  This still leaves open a range of possibilities, and to the extent any of these approximations are valid, the results should not depend on precisely how we choose to parameterize the variance (we give a {\em post facto} justification for the insensitivity to the particular approximations through the analysis in sec.~\ref{sec:analysis} below). We will thus make the ``simple" choice  
$$
\sigma(\{M_{n-1},\ldots,M_0\}) \propto M_{n-1} + M_{n-2} + \cdots + M_{n-{L}}.
$$
for parameterizing the variance in terms of the state. With this choice, we can estimate the weights $h'_k$ by the following Monte-Carlo procedure:\\

\noindent {\bf Algorithm II : Filters determined by statistical regression}
\label{alg2}
\begin{enumerate}
\item Pick initial conditions $\ket{\rho_0^{(j)}}$ for $j=1,2,\ldots,J$ by sampling from an appropriate distribution.
\item For each initial condition $\ket{\rho_0^{(j)}}$, generate the sequence $M_k^{(j)}$ for $k=1,2,\ldots,N$. This is the ``training data set".
\item Find the weights $h'_k$ by minimizing the sum of the normalized squared residuals
$\displaystyle{\sum_{j=1}^J \sum_{n=L+1}^{N} \left[ \frac{M^{(j)}_{n} - \sum_{k=1}^{L} h'_k M^{(j)}_{n-k}}{\sum_{k=1}^{L} M^{(j)}_{n-k}}\right]^2}$, where the outer sum is over different realizations, and the inner sum is over all subsequences of $L$ consecutive values of $M^{(j)}_k$. The resulting equations are of course the analogs of the {\em Yule-Walker} equations \cite{Yule_Method_1927,walker,box-jenkins} for the design of AR filters for stationary random processes. In our case, the covariances $\mathbb{E}[M_n M_k]$ are estimated by averaging over time in each realization and also averaging over an ensemble of realizations. 
\item Once we determine the parameters $h'_k$, we can then determine the parameters $q_n$ by minimizing $\displaystyle{\sum_{j=1}^J  \left[ M^{(j)}_n - \sum_{k=1}^{n} h'_k M^{(j)}_{n-k} - M_0 q_n\right]^2}$, where we assume that the solutions have been scaled such that $M_0$ is the same for the $L$ independent realizations. Note that we are only averaging over different realizations, and not over time, so we do not have the issue of estimating the variance of a non-stationary noise process. Minimizing over the choice of $q_j$ yields
$$
q_n = \frac{1}{JM_0} \sum_{j=1}^J  \left[M^{(j)}_n - \sum_{k=1}^{n} h'_k M^{(j)}_{n-k} \right]
$$
\item An obvious modification of this method also applies to determine the regression coefficients for estimating $M_{n}^{-1}$ from its history. Since $(M+\delta)^{-1} \approx M^{-1} - \delta/M^2$ if $\delta/M \ll 1$, we can postulate the variance parameterization for the filter
$$
\bar{M}^{-1}_{n} = q'_{n} + \sum_{k=1}^{\min(n,L)} \nu'_k M^{-1}_{n-k} + \sigma'(\{M_{n-1},M_{n-2},\ldots,M_0\}) \theta'_n
$$ 
as $\sigma' \propto (M_{n-1} + M_{n-2} + \cdots + M_{n-{L}})/M_{n}^2 \sim M^{-1}_{n-1} + M^{-1}_{n-2} + \cdots + M^{-1}_{n-{L}}$. \\
\qed

\end{enumerate}

  \begin{figure}[tbhp] 
         \centering
         (a) \\
         \includegraphics[width=0.95 \textwidth]{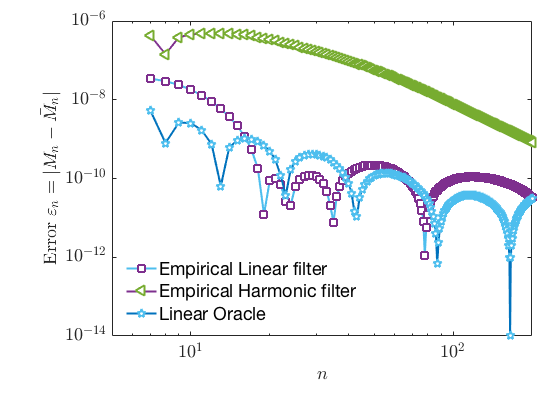} \\
         (b) \\
                    \includegraphics[width=0.95 \textwidth]{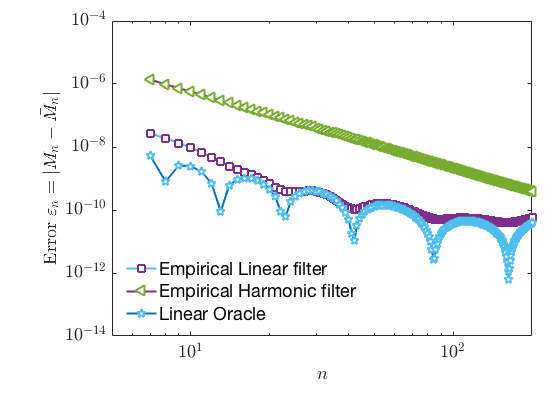}
         \caption{ The performance of the empirical filters determined by regression using many model runs. (a) One-step prediction error for a single realization. The dips in the error correspond to times where the error changes sign, and the number of such changes is naturally related to the number of taps in the filter. (b) The average performance of the empirical filters over 100 random realizations. The empirical linear filter performs better than the empirical harmonic filter for all $n$. It also tracks the average performance of (the realization dependent) linear oracle, except at the dips, which persist even upon averaging. }
         \label{fig:empirical}
      \end{figure}
      
We display the results of this procedure in Fig.~\ref{fig:empirical}. We obtain the empirical AR($L$) estimators $(q_k,h'_k)$ and $(q'_k,\nu'_k)$ with $L = 6$  by averaging the residuals over 100 independent realizations, each starting with $M_0 = 1$ and run for 200 steps. The resulting filters are given by
\begin{align}
\label{eq:bavg}
\bar{M}_n  =  & \ 5.2218 M_{n-1}  - 11.3232 M_{n-2} + 13.0495 M_{n-3} - 8.4286 M_{n-4} \\
& + 2.8926 M_{n-5}  -0.4120 M_{n-6} \nonumber \\
\bar{M}_{n}^{-1} = & \  4.9161 M_{n-1}^{-1} -10.1034 M_{n-2}^{-1} + 11.1366  M_{n-3}^{-1}  - 6.9607 M_{n-4}^{-1}   \label{eq:havg} \\  & + 2.3446 M_{n-5}^{-1}    - 0.3332 M_{n-6}^{-1}  \nonumber 
\end{align}

We test the performance of these filters on 100 new realizations, that were not part of the training set, and also construct, for each realization, a {\em linear oracle}, i.e. a filter that has knowledge of the future, by minimizing the sum of the squared residuals over all subsequences of consecutive values of $M_n$ {\em for this realization}. By construction, the linear oracle has the smallest possible error among all linear filters for the given realization, and is thus a good benchmark for measuring the performance of any given linear filter.

\begin{figure}[tbhp] 
         \centering
        \includegraphics[width=0.95 \textwidth]{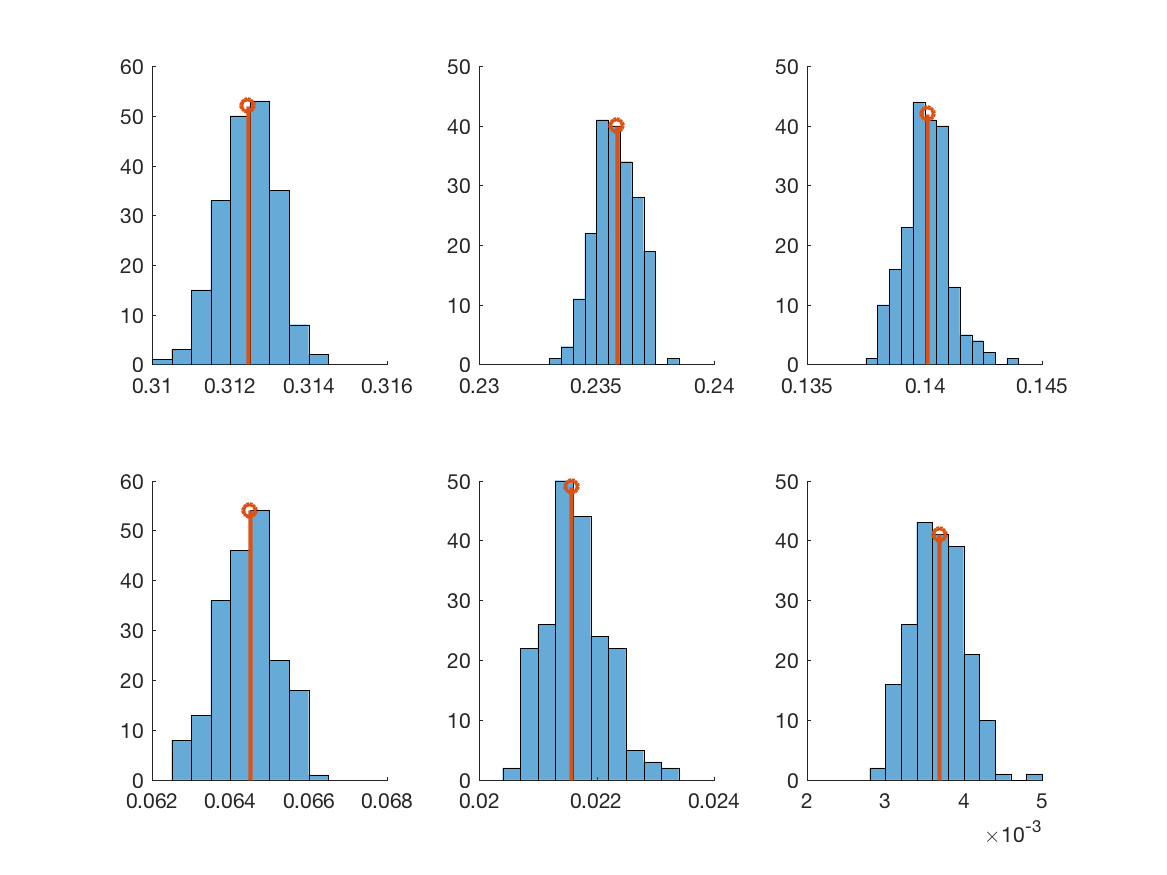}    
         \caption{Distribution of the poles of the linear oracles for 200 independent realizations. The abcissa in each plot is $1-z_j, j =1,2,\ldots,6$ where $z_j$ is the $j$th smallest pole. The empirical filter is obtained by averaging the residuals over time and also the first 100 realizations. It has poles at $0.9963,0.9784,0.9355,0.8599,0.7641$ and  $0.6876$ respectively, and these locations are marked on the corresponding histograms. Note that the relative variation in the pole location, among the various realizations, gets larger, exponentially in $j$, showing that the pole locations are sloppy parameters.  }
         \label{fig:optimal-theory}
      \end{figure}

The process $M_n$ is not stationary, and certainly not ergodic, so there is no reason  to expect that we can replace an ensemble average over different realizations by a time average. Also, as discussed above, the procedure for computing the linear oracle cannot be done `online' (i.e. as the data $M_n$ is being generated)  because we need the entire history of the sequence $M_n$ to compute it. Nonetheless, as we illustrate in fig.~\ref{fig:optimal-theory}, the variations between the linear oracles for different  realizations are small, and they all agree with the filter generated by averaging the residuals over time and realizations. This is also reflected in the fact that the averaged filter performs nearly as well as the linear oracle, for data {\em that were not part of the training set}.

\section{Asymptotic Filters} \label{sec:analysis}

In the previous section we considered various approaches to building a stochastic parametrization of the process $M_n$. The numerics revealed some counter-intuitive results. The numerics also demonstrate, unequivocally, that the `best' stochastic parameterization came from the data-driven empirical linear filter. In this section, we develop a framework for the analyzing the continuum model~\eqref{eq:continuum} in a probabilistic setting with  random initial conditions drawn from a distribution. We also characterize the noise process $\beta_n$ in the Mori-Zwanzig projection~\eqref{mz2}, and this allows us to rigorously analyze the stochastic parametrization/filtering schemes from Section~\ref{sec:parametrization}.  Through this analysis we provide an explanation for the observations from numerical simulations, and also present some new theoretical insights into stochastic parametrization/prediction for systems with slow relaxation/algebraic decay of correlations.

In our numerical discretization of ~\eqref{eq:continuum}, we take
\begin{equation}
\rho^I_0(w) = \sum_{i=1}^I \Delta \gamma_i \delta\left(w - \left(\frac{i}{I} - \frac{1}{2I}\right)\right),
\label{init-cond-sequence}
\end{equation}
where $\Delta = 1/I$ and the quantities $\gamma_i$ are i.i.d non-negative random variables with mean $\mu_\gamma = \mathbb{E}[\gamma]$ and variance $\sigma_\gamma^2 =  \mathbb{E}[(\gamma -\mathbb{E}[\gamma])^2] $. If $\phi,\psi$ are continuous functions on $[0,1]$, direct calculations show that
$$
\mathbb{E}\left[\int_0^1 \rho_0(w) \phi(w) dw\right] =  \mathbb{E}[\gamma]\sum_{i=1}^I \phi \left(\frac{i}{I} - \frac{1}{2I}\right) \Delta \approx \mu_\gamma \int_0^1 \phi(x) dx, 
$$
and
\begin{align*}
& \mathbb{E}\left[\iint \phi(w) \psi(w') \rho_0(w) \rho_0(w') dw dw'\right]  \\ = 
& \left[\mathbb{E}[\gamma]\sum_{i=1}^I \phi \left(\frac{i}{I} - \frac{1}{2I}\right) \Delta\right]^2  + \mathbb{E}[(\gamma-\mathbb{E}[\gamma])^2] \sum_{i=1}^I  \phi \left(\frac{i}{I} - \frac{1}{2I}\right) \psi \left(\frac{i}{I} - \frac{1}{2I}\right) \Delta^2 \\
\approx &\ \  \mu_\gamma^2 \iint \phi(w) \psi(w') dw dw'  +  \frac{\sigma_\gamma^2}{I} \int \phi(w) \psi(w) . 
\end{align*}
The distribution for $\gamma$ can potentially depend on $I$. We will assume that  $\mu_\gamma \to 1$ and $\frac{\sigma_\gamma^2}{I} \to \bar{\sigma}^2$ as $I \to \infty$. If the distribution of the weights $\gamma_I$ is independent of $I$, then either $\bar{\sigma}^2 = 0$, the {\em central limit theorem} scaling, or $\bar{\sigma}^2 = \infty$, and neither situation is appropriate for modeling natural oil where we expect that there is some finite variance associated with the uncertainty  in the composition of oil. Thus, we need to sample from initial conditions  with $0 < \bar{\sigma}^2 < \infty$.

For any prescribed value $0 < \bar{\sigma}^2 < \infty$, we can indeed find a family of $I$-dependent distributions that satisfy these conditions, by appropriately truncating and rescaling a distribution that has finite mean but infinite variance. The details are presented in Appendix~\ref{apndx:weak-limit}.

Under the conditions $\mu_\gamma \to 1$ and $\frac{\sigma_\gamma^2}{I} \to \bar{\sigma}^2$ as $I \to \infty$, we have a distribution of initial conditions $\ket{\rho_0}$ defined by the weak limits of sequences of the form~\eqref{init-cond-sequence}. For any pair of observables $\phi$ and $\psi$, we have 
\begin{align}
\label{moments}
\mathbb{E} [\bra{\phi}\ket{\rho_0}] & = \bra{\phi}\ket{1} \equiv \bra{\phi}\ket{\mathbb{E} [\rho_0]} \\
\mathbb{E} [\bra{\phi}\ket{\rho_0} \bra{\psi}\ket{\rho_0}] & = \bra{\phi}\ket{1}\bra{1}\ket{\psi} + \bar{\sigma}^2 \bra{\phi}\ket{\psi} \equiv \bra{\phi} \mathbb{E}[\ket{\rho_0}\bra{\rho_0} ] \ket{\psi}. \nonumber
\end{align}
These equations characterize the first two moments of the distribution on the initial conditions for the continuum limit~\eqref{eq:continuum}. Alternatively, we could have postulated that the distribution of the initial conditions had these expressions for the first and second moments by requiring that the distribution be invariant under all the (mathematically simplifying but not physically well motivated) exchanges $w \to w'$ that interchanges the amounts of two distinct species in the oil. This naturally leads to $\mathbb{E}\ket{\rho_0} \propto \ket{1}$ and $\mathbb{E}\ketbra{\rho_0}{\rho_0}$ is a linear combination of $\ketbra{1}{1}$ and the identity operator on $\mathcal{H}$ as these are the only invariant operators under the permutation $w \to w'$. The homogeneity of the model~\eqref{eq:continuum} allows us to eliminate one parameter from the distribution of initial conditions by rescaling the initial mass to $M_0=1$. The symmetry argument therefore allows for a two parameter family of distribution of initial conditions $\ket{\rho_0}$ characterized by 
$$
\mathbb{E} [\braket{\phi}{\rho_0}] = \braket{\phi}{1}, \qquad \mathbb{E} [\bra{\phi}\ket{\rho_0} \bra{\psi}\ket{\rho_0}]  = \zeta^2 \braket{\phi}{1}\braket{1}{\psi} + \bar{\sigma}^2 \braket{\phi}{\psi}, \zeta^2 \geq 1.
$$
  In this view, we have given an explicit construction for how to sample initial conditions from a one-parameter subfamily (corresponding to $\zeta^2 = 1$) of such measures, as (subsequential) weak limits of measures consisting of finitely many point masses.

We now characterize the noise process $\beta_n$ in \eqref{mz2}. In the Schr\"{o}dinger picture the mass at time $n$ is given by the observable $\bra{g} = \bra{1}\Lambda$ acting on the state $\ket{\rho_{n-1}}$ at time $n-1$. Using this with \eqref{master-eqn}, we get 
$$
h_k = \mel{1}{\Lambda (Q\Lambda)^k}{1}, \qquad \beta_n = \mel{1}{(\Lambda Q)^n}{\rho_0}.
$$
Consequently,
$$
\mathbb{E}[\beta_n] = \mel{1} {(\Lambda Q)^{n}}{\mathbb{E}[\rho_0]} = \mel{1}{(\Lambda Q)^{n}}{1} = 0,
$$
and
\begin{align*}
\mathbb{E}[\beta_n \beta_m] & = \bra{1}(\Lambda Q)^{n} \mathbb{E}[\ketbra{\rho_o}{\rho_0}] (Q \Lambda )^{m} \ket{1} \\
& = \bra{1}(\Lambda Q)^{n} (P + \bar{\sigma}^2 ) (Q \Lambda )^{m} \ket{1} \\
& = \bar{\sigma}^2 \mel{1}{(\Lambda Q)^{n+m-1} \Lambda}{1} \\
& = \bar{\sigma}^2 h_{n+m-1}.
\end{align*}
We have used $PQ = 0, Q^2 = Q$. This is a fluctuation-dissipation relation for the system in \eqref{eq:continuum}, and not unexpectedly, it {\em does not have the form of the usual Fluctuation-Dissipation theorem} for a Hamiltonian system with short memory \cite{Crisanti_Violation_2003}. Note also, that
\begin{align*}
\mathbb{E}[\beta_n M_j] & = \bra{1}(\Lambda Q)^{n} \mathbb{E}[\ketbra{\rho_o}{\rho_0}] \Lambda^j \ket{1} \\
& = \bra{1}(\Lambda Q)^{n} (P + \bar{\sigma}^2 ) \Lambda ^{j} \ket{1} \\
& = \bar{\sigma}^2 \mel{1}{(\Lambda Q)^{n} \Lambda^j}{1}.
\end{align*}
These expectations are non-zero in general; we compute them explicitly in Appendix~\ref{sec:orthogonal}.  

This result is somewhat unexpected. It is certainly true that $\braket{F_n}{\xi_n} = \mel{F_n}{P \Lambda^n}{\rho_0}  = 0$ (see discussion before \eqref{orthogonal}) for any initial measure $\rho_0$,  but this does not imply that $\mathbb{E}[\beta_n M_j] = \mathbb{E}[\braket{F_n}{\rho_0} \braket{\rho_0}{\xi_n}] = 0$. The issue is that the observable $g^t$ given by the constant function $g^0(x) = 1$ has non-trivial evolution (see discussion after \eqref{heisenberg}) in contrast to the situation for dynamical systems, so the usual intuition does not apply.  The ``noise" is no longer uncorrelated with the observations, and  this explains why the projection formalism {\em does not give an optimal prediction/stochastic parametrization procedure} for~\eqref{eq:continuum}.  The empirical  filters, albeit still linear, and with far fewer taps, do perform better than the MZ estimator, because, by construction, the ensemble average of the product of the noise and  past observations is zero. 

\subsection{Optimal time-varying filters} \label{sec:asymptotic}

We now address the question of why the empirical linear filter performed almost as well as the linear oracles. For the process $M_n = \mel{1}{\Lambda^n}{\rho_0}$, we have 
$$
\mathbb{E}[M_n] = \bra{1}{\Lambda^n}\mathbb{E}[\ket{\rho_0}] = \mel{1}{\Lambda^n}{1} = \frac{1-e^{-n\tau}}{n \tau},
$$ 
and  
$$\mathbb{E}[M_n M_j]  = \mel{1}{\Lambda^m \mathbb{E}[\ketbra{\rho_0}{\rho_0}] \Lambda^j}{1} = \mathbb{E}[M_n]  \mathbb{E}[M_j] + \bar{\sigma}^2 \mel{1}{\Lambda^{n+j}}{1}. 
$$
By regression, there is indeed an optimal AR($L$) filter of the form 
$$
M_{n} = q_{n} M_0 + h_1^{(n)} M_{n-1} + h_2^{(n)} M_{n-2} + \cdots  + h_{L}^{(n)} M_{n-L} + \theta_n,
$$
where the innovation $\theta_n$ is orthogonal to  $M_{n-1},M_{n-2},\ldots,M_{n-L}$ and $M_0 = 1$. Note that the optimal filter is allowed to (and as we see below, in general does) depend on $n$, so it is not autonomous. The orthogonality condition gives the Yule-Walker equations
\begin{align}
\label{yule-walker}
\mathbb{E}[M_{n}M_{n-k}]  & = q_{n} \mathbb{E}[M_{n-k}] + \sum_{j=1}^{L}  h_j^{(n)} \mathbb{E}[M_{n-j}M_{n-k}], \\
\mathbb{E}[M_{n}]  & = q_{n} + \sum_{j=1}^{L}  h_j^{(n)} \mathbb{E}[M_{n-j}]. 
\end{align}
Multiplying the second row by $\mathbb{E}[M_{n-k}]$ and subtracting from the first row gives
$$
\frac{1 -e^{(2 n- k) \tau}}{(2n-k)\tau} = \sum_{j=1}^{L} h^{(n)}_j  \frac{1 -e^{(2 n- k -j) \tau}}{(2n-k-j)\tau}, \quad k = 1,2,\ldots,L.
$$
Note that these equations are independent of $\bar{\sigma}^2$, so they do not require small noise. For $n - L \gg \tau^{-1} \sim O(1)$, we can ignore the exponentially small quantities and get the matrix system  $v= Ah$ where the vector $v$ and matrix $A$ are as defined below,
\begin{equation}
\begin{pmatrix} \frac{1}{2n-1} \\ \frac{1}{2n-2} \\ \vdots \\   \frac{1}{2n-L}\end{pmatrix} = \begin{pmatrix} \frac{1}{2n-2} & \frac{1}{2n-3} & \cdots &  \frac{1}{2n-L-1} \\
 \frac{1}{2n-3} & \frac{1}{2n-4} & \cdots &  \frac{1}{2n-L-2} \\
\vdots & \vdots & \ddots & \vdots \\
\frac{1}{2n-L-1} & \frac{1}{2n-L-2} & \cdots &  \frac{1}{2n-2L} \end{pmatrix}
\begin{pmatrix} h_1^{(n)} \\ h_1^{(n)}  \\ \vdots \\   h_{L}^{(n)} \end{pmatrix}.
\label{eq:asymp-matrix}
\end{equation}
The coefficient matrix $A$ for this system is a variant of the classical {\em Hilbert matrix}, a well known example of an ill-conditioned matrix. The condition number of this matrix is $\sim (2n)^{L}/L!$ which can be enormous, and we cannot solve the system in a numerically stable manner, although, of course, a unique solution does exist. We compute solutions to this system in  Appendix~\ref{sec:cauchy} to obtain
\begin{align}
h^{(n)}_j & = \prod_{i \neq j}^L \frac{i}{i-j} \prod_{i=1}^L \frac{2n - i -j}{2n - i} \label{eq:asymp-solution} \\
&  = (-1)^{j-1}\binom{L}{j} + (-1)^j \frac{L^2}{2n} \binom{L-1}{j-1} + O(n^{-2}). \nonumber
\end{align}

We can  fix $L$, the order of the filter, and look at the behavior of the filter coefficients as expansions in $n$. Picking $L  =6$ (corresponding to predicting $M_{n}$ using $M_{n-1},M_{n-2},\ldots,M_{n-6}$) for illustration, we get
\begin{align}
\label{asymp-coeffs}
h^{(n)}_1 & = 6 - \frac{36}{2n-1}, \\
h^{(n)}_2 & =-15+\frac{630}{2 n-1}-\frac{225}{n-1}, \nonumber \\
h^{(n)}_3 & = 20-\frac{3360}{2 n-1}+\frac{2100}{n-1}-\frac{1200}{2 n-3},\nonumber \\
h^{(n)}_4 & =-15 +\frac{7560}{2 n-1}-\frac{6300}{n-1}+\frac{6300}{2 n-3}-\frac{450}{n-2},\nonumber \\
h^{(n)}_5 & = 6 -\frac{7560}{2n-1}+ \frac{7560}{n-1}-\frac{10080}{2 n-3}+\frac{1260}{n-2}-\frac{180}{2 n-5}, \nonumber \\
h^{(n)}_6 & =-1+\frac{2772}{2n-1}-\frac{3150}{n-1}+\frac{5040}{2 n-3}-\frac{840}{n-2}+\frac{210}{2 n-5}-\frac{3}{n-3}. \nonumber 
\end{align}
The (non-autonomous) filter 
\begin{equation}
\bar{M}_n = \sum_{j = 1}^6 h^{(n)}_j M_{n-j}
\label{asymp-filter}
\end{equation}
with the coefficients given by \eqref{asymp-coeffs} is the {\em Asymptotic filter} with 6 taps. Clearly,  the filter coefficients converge as $n \to \infty$
$$
\lim_{n \to \infty} h^{(n)}_j = (-1)^{j-1} \binom{L}{j}
$$
This is a {\em post facto} justification for why we could average over $n$, in addition to averaging over independent ensembles, in determining the coefficients of the empirical filter (Section~\ref{sec:empirical}).

\subsection{Universal filters for slowly decaying correlations}

There is a satisfying intuitive explanation for the form of the asymptotic limit filter. Consider the problem of finding coefficients $\alpha_0,\alpha_1,\alpha_2,\ldots,\alpha_{L}$ such that the asymptotic growth for the linear combination 
$
\sum_{i=0}^L \alpha_i (n - i)^{-1} 
$
is as small as possible, where we normalize the coefficients by requiring that $\alpha_0 = 1$. It is clear that for generic choices of $\alpha_i$, the decay rate of the combination is $O(n^{-1})$, but we can do better by judicious choices of $\alpha$. For example, $\alpha_0 = 1, \alpha_1 = -1$ and the rest of the $\alpha_i = 0$ gives a decay rate $O(n^{-2})$. The smallest possible asymptotic behavior comes from the coefficients $\alpha_i$ set equal to a row of Pascal's triangle with alternating signs, as we can see from an inductive argument. For this choice of $\alpha_i$, we have
$$
\sum_{i=0}^L \binom{L}{i} \frac{(-1)^i}{n-i} = \frac{L!}{n(n-1)(n-2)\cdots(n-L)} \sim \frac{L!}{n^{L+1}},
$$  
and for any other choice of the coefficients, the decay of the linear combination is slower.
 So it is indeed to be expected that if the optimal filter coefficients converge 
 $\lim_{n \to \infty} h_j^{(n)} = \alpha_j$, then $\alpha_j = (-1)^j \binom{L}{j}$.
The coefficient $q_{n}$ is determined by \eqref{yule-walker} as 
$$
q_{n} = \mathbb{E}[M_{n}]  - \sum_{j=1}^{L}  h_j^{(n)} \mathbb{E}[M_{n-j}]  \sim \frac{L!}{n^{L + 1}} + o(n^{-L - 1}).
$$
 Thus $q_{n}$ decays very rapidly so that it can be set to zero for $n > L \sim O(1)$.

These filter coefficients are `universal' for all processes with slowly decaying correlations. Indeed, for a slowly decaying function $f(x)$ (say one consisting of nonpositive powers of $\log(x)$ and of $x$), we have
$$
\sum_{i=0}^L \binom{L}{i} (-1)^i f(n-i) \sim \frac{d^L}{dx^L} f\left(n-\frac{L}{2}\right),
$$  
and one cannot get better asymptotic decay with constant coefficient linear combinations of $L+1$ consecutive terms. It is thus tempting to suggest that all  processes with slowly decaying correlations can be stochastically parameterized by
\begin{equation}
[(1-R)^L f]_{n} = \sum_{i=0}^{L} \binom{L}{i} (-1)^{i} f_{n-i} = \sigma_n \theta_n,
\label{state-space}
\end{equation}
where $R$ is the right shift operator on sequences, $[R f]_n = f_{n-1}, \theta_n$ are independent normal variates and the variance parameter $\sigma_n$ has statistics that can be estimated from  data. Eq.~\eqref{state-space} is the model reduction that is associated with the filter
\begin{equation}
\label{universal}
\bar{M}_n =  \sum_{j=1}^L (-1)^{j-1} \binom{L}{j} M_{n-j},
\end{equation}
which is an {\em universal} filter for processes with slowly decaying correlations, since it is expected to work just as well for any such process.

The universal stochastic parametrization \eqref{state-space} does not depend on the correlation structure of the process that is being modeled, besides requiring that it decay algebraically.  Thus one does not expect this model to track a realization of the underlying process without additional data assimilation. Nor does one expect that an ensemble of solutions of~\eqref{state-space} with appropriate statistics for $\theta_n$ necessarily reproduce the statistics of an ensemble of realizations of the underlying process.  Indeed, the transfer function of the universal filter is $(1-z^{-1})^{-L}$ and has a pole of order $L$  at $z = 1$. The filter is thus unstable, and has homogenous solutions $f_n = n^j$ for $j=0,1,2,\ldots,L-1$ which do not decay to 0.

\begin{figure}[tbhp] 
         \centering
         \includegraphics[width=0.95 \textwidth]{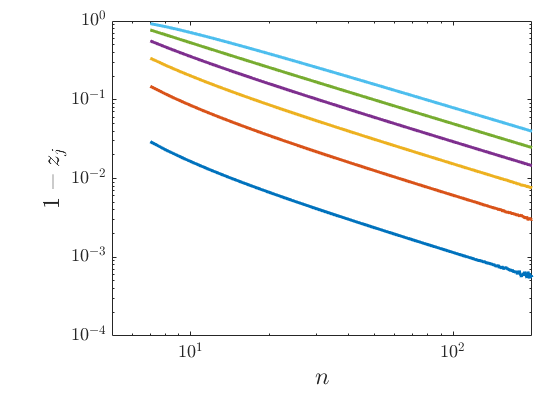}    
         \caption{The poles $z_j, j = 1,2,\ldots, L$ of the asymptotic filter~\eqref{asymp-filter} as a function of $n$ for $L = 6$. To show the convergence to 1 we plot $1-z_j \approx -\log(z_j)$, and these quantities are all positive indicating that the filters are stable. Also, for each of the poles, we see that $z_j  \approx 1-c_j n^{-1}$ from the slope of the corresponding graph. These curves also demonstrate that our model is sloppy, as evident from the level repulsion between the poles $z_j$ at every fixed value of $n$. }
         \label{fig:poles}
      \end{figure}

We can attempt to remedy these shortcomings by going to higher order in the solutions of the Yule-Walker equations \eqref{asymp-filter}. The $n$-dependent corrections to the limiting filter coefficients do reflect the particular correlation function $\mathbb{E}[M_n M_j] \sim 1/(m+j)$ for the evaporation process and are thus {\em not universal}. The correction depends explicitly on $n$ so including these corrections will make the filter non-autonomous. However, these corrections make the filter stable. Figure~\ref{fig:poles} shows the poles of the filter transfer function 
$$
H^{(n)}(z) = \frac{1}{1- \sum_{j=1}^L h_j^{(n)} z^{-j-1}},
$$
corresponding to a shift invariant filter obtained by ``freezing" the time index $n$. Note that all the poles are real, less than 1, and approach $1$  as $n \to \infty$ from inside the unit circle.

If we order the poles by $z_j \leq z_{j+1}$, we have $z_j \approx 1 - c_j n^{-1}$, and the constants $c_j$ are (roughly) geometrically distributed, i.e $\log(c_{j-1}/c_j) \sim O(1)$ (strong level repulsion) for all $j$, indicating that the poles are sloppy parameters \cite{Waterfall_Sloppy_2006} for the description of the linear evaporation process~\eqref{eq:continuum}. The filter coefficients $h^{(n)}_k$ are determined by
$$
1- \sum_{k=1}^L h_k^{(n)} z^{-j-1} =   \prod_{j=1}^L \left(1 - \frac{z_j}{z}\right)
\approx \prod_{j=1}^L \left(1 - \frac{1 - c_j n^{-1}}{z}\right)
$$
so that the filter coefficients are $n$ dependent, {\em symmetric} functions of the quantities $c_j$. Although the poles are sloppy, the filter coefficients themselves are robust \cite{Waterfall_Sloppy_2006}. We first learned this principle, {\em viz.} symmetric functions of random quantities are computable in terms of the low order moments of their distribution, and are hence robust, in work with Leo Kadanoff on the extremal distribution of points for the Thomson problem in 2D domains \cite{Berkenbusch_Discrete_2004}.

\begin{figure}[tbhp] 
         \centering
         \includegraphics[width=0.95 \textwidth]{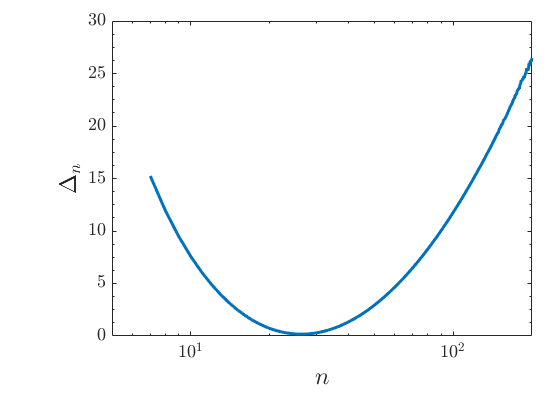}    
         \caption{The discrepancy $\Delta_n$ between the poles of the empirical filter and the poles of the asymptotic filter~\eqref{asymp-filter}. The empirical filter was constructed as described in sec.~\ref{sec:empirical} by running the process for $N=200$ time steps and averaging the residuals over 100 realizations. The poles of the empirical filter are at $0.9963,0.9784,0.9355,0.8599,0.7641$ and  $0.6876$ respectively. }
         \label{fig:discrepancy}
      \end{figure}

We can now given an analytical explanation for the reason that the empirical linear filter performed almost as well as the genie-aided linear oracle, and thus is demonstrably a near-optimal linear filter, among all shift-invariant linear filters with $L$ (a given number of) taps.  For an interval of time $1 \leq k \leq N$, we can pick an autonomous filter that is (approximately) optimal for the entire range by using the corrections in \eqref{asymp-filter} with $n = \bar{\kappa}$ being an appropriate ``averaged" time index over the interval of interest. This will give a filter with fixed coefficients and $L$ taps, that is guaranteed to be stable. Since the corrections in \eqref{asymp-filter} are $O(n^{-1})$ it is not unreasonable to expect that  
$$
\frac{1}{\bar{\kappa}} \sim \frac{1}{N} \sum_{j=L}^N \frac{1}{j} \sim \frac{\log(N)}{N},
$$
so that $\bar{\kappa}$ is the harmonic mean of the time interval. For $N = 200$, we would estimate $\bar{\kappa} \approx 37.75$. Since the quantity of interest is the deviation of the $j$th pole from 1. We define the discrepancy
$$
\Delta_n = \sum_{j =1}^L \left|\log\left(\frac{1-z_j^{(n)}}{1 - \tilde{z}_j}\right) \right|^2,
$$
where $z_j^{(n)}$ is the $j$-th pole of the asymptotic filter~\eqref{asymp-filter} (shown in fig.~\ref{fig:poles}) and $\tilde{z}_j$ is the $j$-th pole of the empirical filter (see fig.~\ref{fig:optimal-theory}). Figure~\ref{fig:discrepancy} shows the discrepancy $\Delta_n$ as a function of $n$. Minimizing the discrepancy, we would infer that $\bar{\kappa} \approx 26.43$ which is on the same scale, although a little smaller than our estimate of $37.75$. 

\section{Multilayer stochastic models for the evaporation of oil} \label{sec:MSM}

The asymptotic filter~\eqref{asymp-filter} is not autonomous, and building a reduced model using this filter will similarly give a non-autonomous stochastic parametrization. We can use a standard `trick' to recast time-dependent systems as autonomous systems on a larger phase space \cite{ott-book}. In particular, by enlarging our ``phase-space" to include an additional dynamical variable $\kappa$  that tracks $n$, we have the autonomous stochastic parametrization
\begin{align}
\label{extended}
M_{n} - \sum_1^L h_j^{(\kappa_n)}M_{n-j} & = \sigma_n \theta_n, \\
\kappa_{n} & = \kappa_{n-1} + 1 + s_n \theta'_n, \nonumber \\
\sigma_n^2 & \simeq \bar{\sigma}^2\frac{L!}{\kappa_n^{L+1}},  \nonumber
\end{align}
where $\theta_n,\theta'_n$ are i.i.d process of normal variates, the dynamical variable $\kappa$ tracks the ``microstructure" of the oil composition in terms of its ``age," and the filter coefficients $h^{(\kappa)}_j$ are given by \eqref{eq:asymp-solution} with $n = \kappa$. 

The reduced model in~\eqref{extended} has the form on a multilayer stochastic model (MSM) \cite{Chekroun_Predicting_2011,Majda_Physics_2013,Kondrashov_Data_2015}, where the quantity $M_n$ is directly observable and the quantity $\kappa_n$ is hidden. To use this model for stochastic parametrization, we can specify $\bar{\sigma}$ and set $s_n = 0$ so that $\kappa_n = n$. Alternatively, given noisy measurements $\tilde{M}_n$, we can estimate the state $(M_n,\kappa_n)$ and a parameter $\bar{\sigma}^2$ jointly by a nonlinear filtering algorithm, for example by the extended Kalman filter \cite{jazwinski-book}. We will present these results in a later publication. 

A crude version of this nonlinear filtering approach is through approximating $\kappa_n$ in terms of $M_{n-1},M_{n-2},\ldots$. In conjunction with~\eqref{extended} and \eqref{eq:asymp-solution}, this will give a stochastic parametrization for $M_n$ without additional `hidden' variables. Since $\mathbb{E}[M_k] \approx \frac{M_0}{k \tau}$, we can define an ``instantaneous age " $\mu_n$ by 
$$
 \frac{1}{\mu_n - 1} \equiv 1 - \frac{M_{n-1}}{M_{n-2}} \approx 1 - \frac{\mathbb{E}[M_{n-1}]}{\mathbb{E}[M_{n-2}]} = \frac{1}{n-1},
$$
so that $\mu_n$ gives an estimate of $n$ based on the ``recent past" $M_{n-1}$ and $M_{n-2}$.
The instantaneous age is a fluctuating quantity, and we can estimate the age $\kappa_n$ by smoothing $\mu_n$ through $\kappa_{n} = (1 - \delta) (\kappa_{n-1}+1) + \delta \mu_n$,  
which is the appropriate filtering strategy for the quantity $\kappa_n$ evolving as in~\eqref{extended} (the `model'), with fluctuating estimates given by $\mu_n$ (the `measurements`). 

\begin{figure}[htbp] 
         \centering
         \includegraphics[width=0.95 \textwidth]{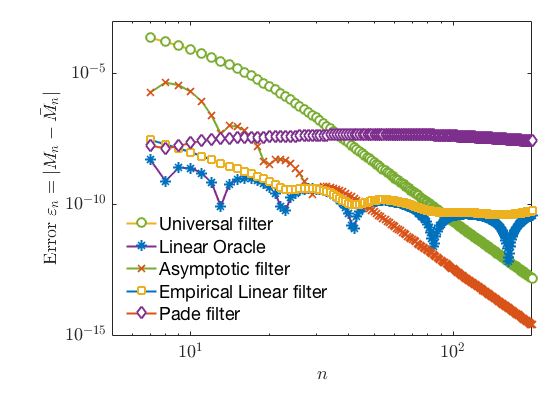}    
         \caption{Comparison between the one-step prediction errors for the the reduced  models given by the  empirical and the asymptotic filters. The empirical filter is generated by averaging the squared residuals over 100 realizations, and the performance of each filter is assessed by averaging errors over another set of 100 realizations. For large $n$, the one-step prediction error of the universal filter \eqref{state-space} is small; nonetheless it is unstable, and cannot be used without incorporating additional information from measurements. The extended asymptotic filter is run with $\delta = 1$ and the corresponding curve is indicated as `Asymptotic filter' in the legend.}
         \label{fig:comparison-grand}
      \end{figure}

Using $\kappa_n$ in place of for $n$ in~\eqref{asymp-filter} gives a nonlinear filtering algorithm for $M_n$, that we will call  an  {\em extended asymptotic filter}. It is a nonlinear modification of~\eqref{asymp-filter} that makes it autonomous. Explicitly, the extended asymptotic filter with $L$-taps is given by $\kappa_0 = 0$ and
\begin{align}
\label{eaf}
\mu_n & = \frac{M_{n-1}}{M_{n-2}-M_{n-1}} + 1, \\
\kappa_n & =  \kappa_{n-1}+1 + \delta(\mu_n -\kappa_n-1) \nonumber\\
h^{(\kappa_n)}_j & = \prod_{i \neq j}^L \frac{i}{i-j} \prod_{i=1}^L \frac{2 \kappa_n - i -j}{2 \kappa_n - i}, \quad j=1,2,3,\ldots,L \nonumber \\
\bar{M}_n & = \sum_{j=1}^L h_j^{(\kappa_n)} M_{n-j} \nonumber
\end{align} 
This filter is  non-empirical, nonlinear, autonomous, independent of realization and is not genie-aided (does not require knowledge of the future). It contains a parameter $\delta$ that we can set, with $\delta = 0$ corresponding to a very stable, but non-responsive filter, while $\delta = 1$ is a very responsive, but potentially numerically unstable filter. We compare the performance of the extended asymptotic filter~\eqref{eaf}  with $\delta =1$, the universal filter~\eqref{universal} with the empirical filters from~\ref{sec:empirical} and the Pad\'{e} filter from~\ref{sec:linear}, when applied to synthetic data. The results are shown in fig.~\ref{fig:comparison-grand}.

Outside of an initial transient, the extended asymptotic filter~\eqref{eaf} is clearly better than the competing methods. The one-step prediction error has the optimal scaling $\varepsilon_n \sim n^{-L-1}$ and by the arguments from before, we would expect this to be the smallest possible scale for the error from a filter with $L$ taps. 

To illustrate the practical application of the extended asymptotic filter, we apply it to empirical evaporation curves for various types of crude oils. Fingas \cite{Fingas_Modeling_2004} has measured the evaporation curves for about 200 different oils (Crudes, Fuel oils, Diesels, etc.)  under a variety of conditions and found that the important parameters are the time of evaporation and the ambient temperature. The time and temperature dependence of the evaporation curves are best fit by one of the following two equations (Eqs.~(10)~and~(11) in \cite{Fingas_Modeling_2004}):
\begin{align*}
\%E & = (0.165(\%D) + 0.045(T-15) ) \log(t)  \quad \mbox{and} \\   
\%E & = (0.0254(\%D) + 0.01(T-15) ) \sqrt{t}
\end{align*}
for oils that follow a ``logarithmic" (respectively ``square-root") equation where $\%E$ is the percentage of oil evaporated at time $t$ in minutes, $\%D$ is the percentage (by weight) of the crude oil that is distilled at $180 \degree$C and $T$ is the ambient temperature in degrees Celsius. We can convert the empirical evaporation curve into the total remaining mass $M(t)$  by 
$$
M(t) = 1 - \frac{\%E}{100}.
$$
The fitting functions (the logarithmic and square-root equations) are clearly not valid if $t$ is too small, since $\%E$ and/or its time derivative blows up as $t \to 0$. They are also not valid for very large $t$ as $\%E$ cannot be more than a 100\%. Finally, they are in arbitrary ``empirical" units. We will non-dimensionalize, as in sec.~\ref{sec:model1} by the (unknown!) evaporation rate of the most volatile component in the oil, and also modify (regularize) the small time behavior of the functions to ensure that $M(0) = 1, \frac{d}{dt}M(0) < \infty$. The regularized, nondimensional functions $M(t)$ are thus in one of two forms:
\begin{align}
M(t) & = 1 - a \log(1+t/t_0) \quad \mbox{and} \label{log}\\
M(t) & = 1 - a(\sqrt{1+ t/t_0}-1) \label{sqrt}
\end{align}
where $t$, $t_0$ (small scale cutoff) and $a \ll 1$ are all dimensionless. These equations necessarily have a limited range of validity since we need $M(t) \geq 0$ for all $t$. Solving eqs.~\eqref{log}~and~\eqref{sqrt} for $M(T_{max}) = 0$, we estimate the ranges of validity by $T_{max} \sim t_0 e^{1/a}$ for the logarithmic equation and $T_{max} \sim t_0/a^2$ for the square-root equation.

\begin{figure}[htbp] 
         \centering
         \includegraphics[width=0.95 \textwidth]{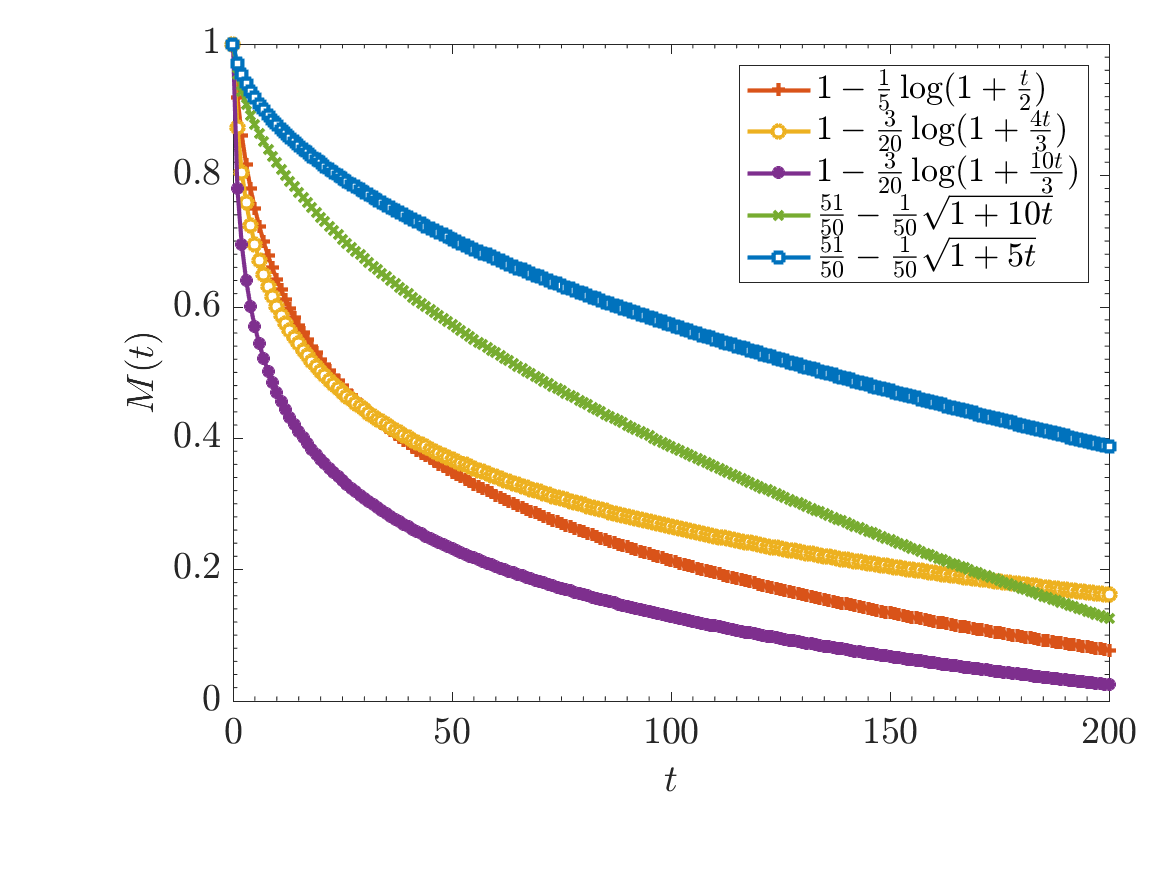}    
         \caption{ The evaporation curves generated by the logarithmic and square-root equations~\eqref{log}~and~\eqref{sqrt}.  The parameters/equations for the curves are given in Table~\ref{tab:parameters}  }
         \label{fig:truth}
      \end{figure}

The parameters $a$ and $t_0$ are related by the following argument. Since we nondimensionalize time by the evaporation rate $\alpha_{max}$ of the most volatile species, the evaporation rate should satisfy  
$$
-\dot{M}(t)  \lesssim \alpha_{max} M(t) = M(t).
$$
Using this inequality for \eqref{log}~and~\eqref{sqrt} at $t=0$ gives $a/t_0 \lesssim 1$. We can also compute $\frac{d^2}{dt^2} \log(M(t))$ for the linear evaporation process~\eqref{eq:continuum} to obtain
$$
\frac{d^2}{dt^2} \log(M(t)) = \frac{ \int w^2 \rho(w,t) dw \int \rho(w,t) dw - \left(\int w \rho(w,t) dw\right)^2}{\left(\int \rho(w,t) dw\right)^2} \geq 0
$$
by the Cauchy-Schwarz inequality. Note that, this relation has to hold for {\em every realization}, and not just in an averaged sense. This relation {\em does not hold} for the logarithmic equation~\eqref{log} for  $t \geq T_{crit} \approx  \frac{1}{e} T_{max}$, although by the time it breaks down, we have $M(T_{crit}) \approx a \ll 1$. 

For the square-root equation~\eqref{sqrt}, the relation fails to hold for $t \geq T_{crit}  \approx  \frac{1}{4} T_{max}$. The empirical square-root fit should therefore break down {\em well before} the total mass hits zero. Indeed, this should already occur by the time $\%E \approx 50\%$. This is an experimentally testable prediction, and checking this will help assess the validity of modeling  assumptions that lead to~\eqref{eq:continuum}. On the other hand, assuming \eqref{eq:continuum} is a good microscopic model, the empirical fits in~\eqref{log}~and~\eqref{sqrt}  can not good models for the ``truth" unless $t \lesssim T_{crit}$.
Consequently, the ability of a filter to track/predict these functions accurately is not necessarily a positive feature. Rather, we would hope that the filters ``discover" that beyond a certain point, the assumed ``truth" actually is not.

\begin{table}[thbp]
   \centering
   \begin{tabular}{@{} cccccr @{}} 
      \toprule
   $T_{max}$ & $a/t_0$  & $a$ & $t_0$ & $T_{crit}$ & Model equation \\
      \midrule
      \\
      \multicolumn{4}{l}{Logarithmic evaporation curve} \\
      \cmidrule(r){1-4} 
      \\
        250 & $0.1$ & $0.2$ & $2$ &  $ \sim 107$ & $M(t) = 1 - \frac{1}{5}\log(1+ \frac{t}{2})$\\
        \\
      	500 & $0.2$ & $0.15$ & $0.75$ &  $\sim 216$ & $M(t) = 1 - \frac{3}{20}\log(1+ \frac{4t}{3})$\\
	\\
	200 & $0.5$ & $0.15$ & $0.3$ &  $\sim 87 $ & $M(t) = 1 - \frac{3}{20}\log(1+ \frac{10t}{3})$ \\
      \midrule
      \\
      \multicolumn{4}{l}{Square-root evaporation curve} \\
      \cmidrule(r){1-4} 
      \\
      250 & $0.2$ & $0.02$ & $0.1$ & $\sim 65$ & $M(t) = \frac{51}{50} - \frac{1}{50}\sqrt{1+ 10t}$\\
      \\
      500 & $0.1$ & $0.02$ & $0.2$ & $ \sim 130$  & $M(t) = \frac{51}{50} - \frac{1}{50}\sqrt{1+ 5 t}$ \\
      \bottomrule
   \end{tabular}
   \caption{Parameter values, model equations and limits on the range of validity  for the numerically generated evaporation curves.}
   \label{tab:parameters}
\end{table}

A final point relates to the role of sampling. In our analysis, $M_n$ is given by $M_n = M(n \tau)$ corresponding to a sampling interval $T_{sampling} = \tau/\alpha_{max}$. In practice, the sampling time is determined by experimental/technical considerations and cannot be freely specified. While $\tau \sim O(1)$ is the ideal situation, so that the time-series data resolves the  dynamics on the fastest scales in the problem, the practically achievable value of $\tau$ can be ``large" and will therefore introduce an additional nondimensional parameter in the discrete time problem. The extended asymptotic filter~\eqref{eaf} and the universal filter~\eqref{universal}, however, are independent of $\tau$. This is a very desirable feature since it allows us to use the same filter independent of the sampling interval. 

To assess the performance of the filters on ``real" data, we use the following numerical procedure: \\

\begin{figure}[tbhp] 
         \centering
         \includegraphics[width= 0.8 \textwidth]{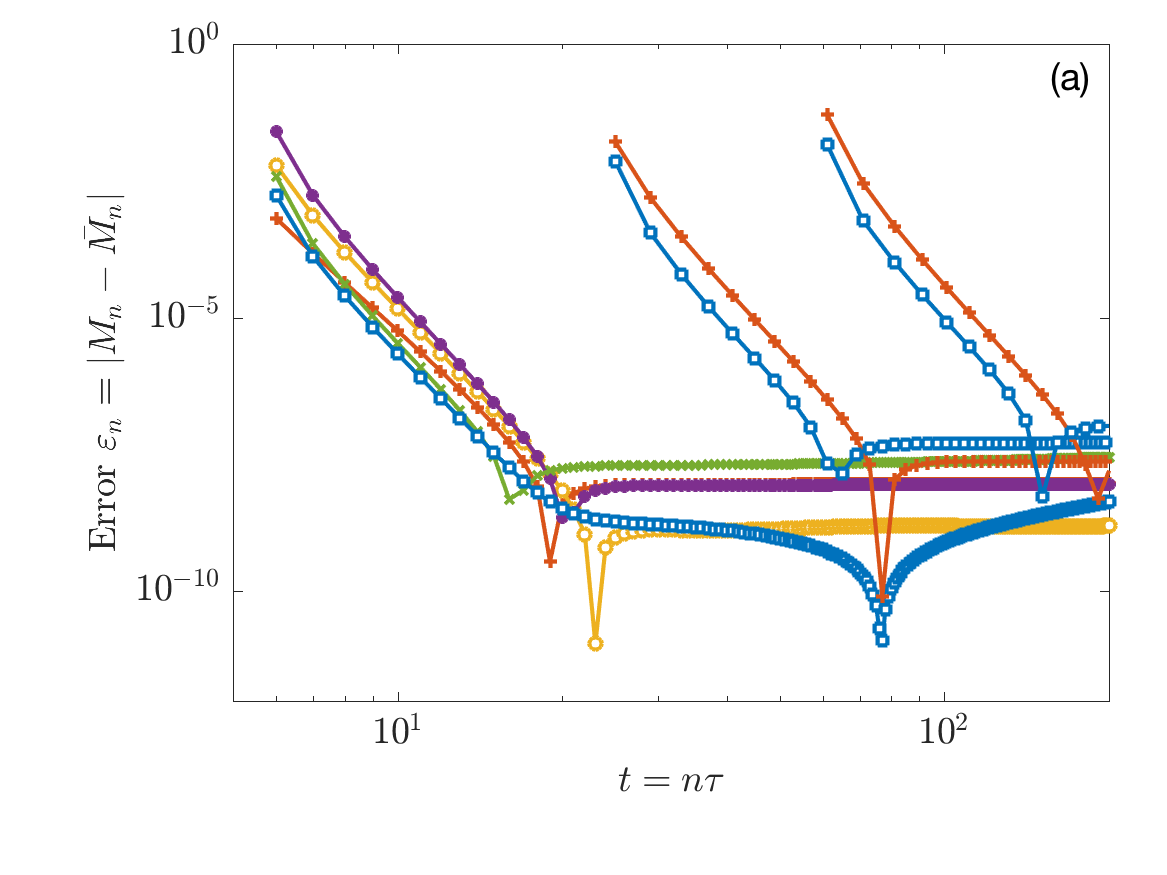}   \\
          \includegraphics[width=0.8 \textwidth]{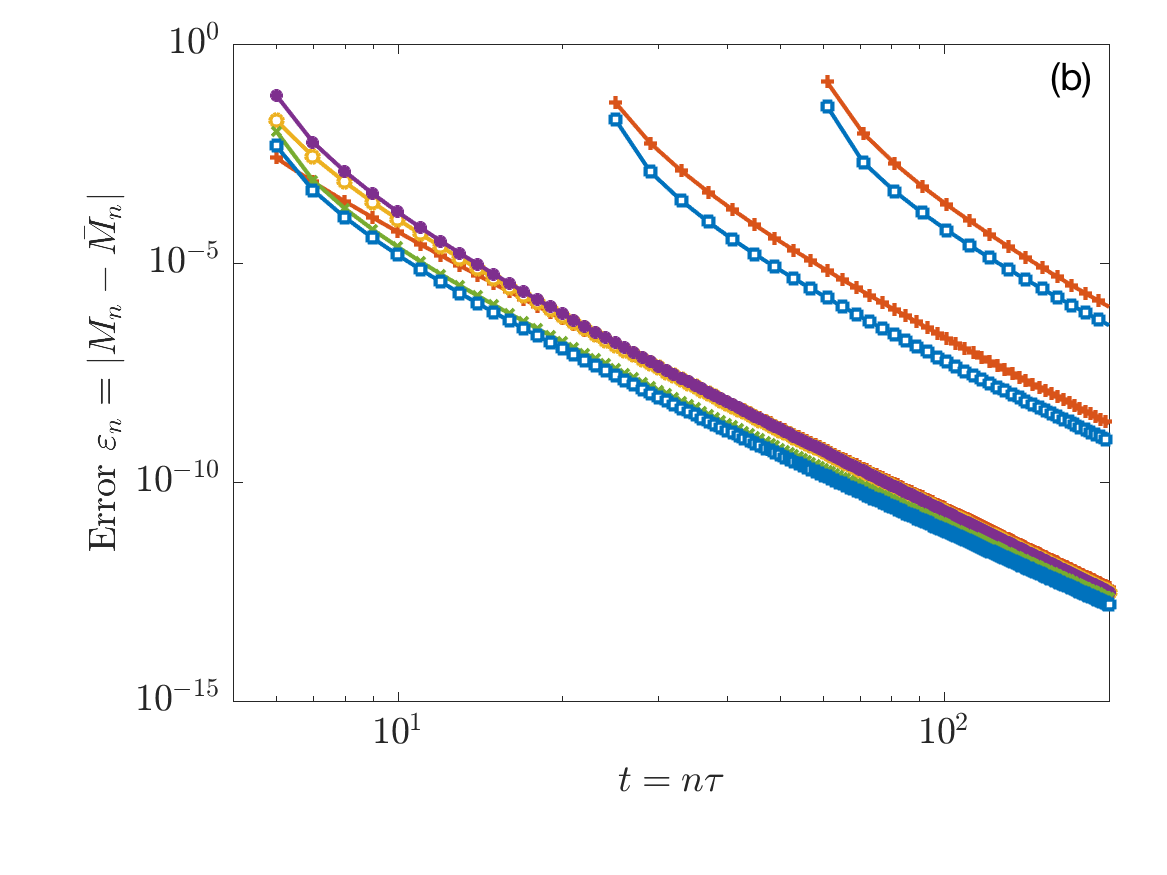}   \\
           \includegraphics[width=0.8 \textwidth]{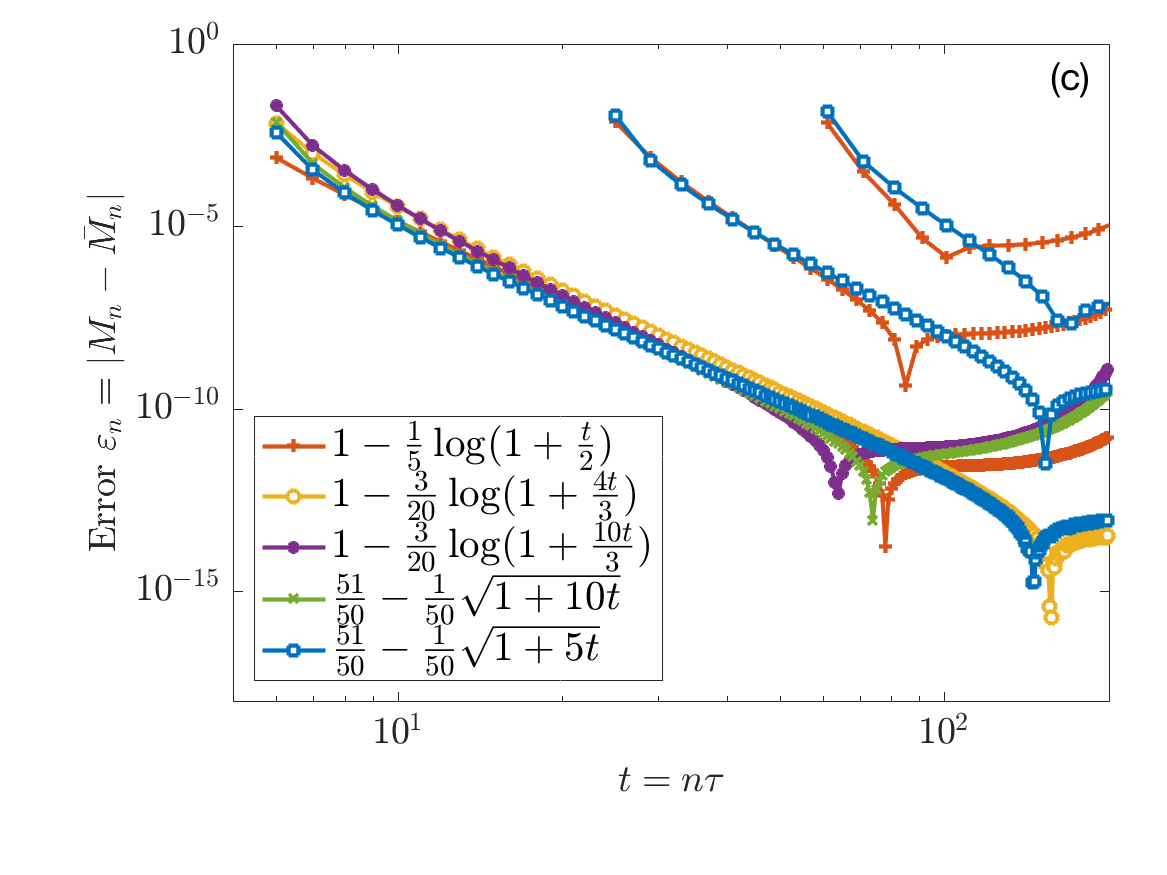}   
         \caption{ One step prediction errors. The filters are applied to time series obtained from the functions in Table~\ref{tab:parameters}, sampled at intervals $\tau$ of 1,4 and 10 time units. (a) Empirical linear filter. (b) Universal filter. (c) Extended asymptotic filter with $\delta = 1$. Note the difference in the vertical scales.}
         \label{fig:comparison-small}
      \end{figure}

\noindent {\bf Algorithm III : Data assimilation and filtering/prediction}
\label{alg3}
\begin{enumerate}
\item Pick parameters $a$ and $t_0$ such that $a/t_0 \lesssim 1$ and $T_{max} \gtrsim 200$. The  parameter values we use are listed in table~\ref{tab:parameters}. 
\item Generate time series $M_n$ using the various parameter values for $a$ and $t_0$ in \eqref{log}~and~\eqref{sqrt} with $0 \leq t = n \tau \leq 200$, where the sampling interval is $\tau = 1,4$ or 10. This procedure corresponds to sampling the curves in Fig.~\ref{fig:truth} at equally spaced intervals. Depending on the sampling rate we get between 20 and 200 samples for each curve.
\item For various time series (different functions, parameters and sampling rates), we compute the one step prediction error for the empirical linear filter~\eqref{eq:bavg}, the universal filter~\eqref{universal} and  the extended asymptotic filter~\eqref{eaf} with $\delta = 1$. These results are shown in Fig.~\ref{fig:comparison-small}.  \qed
\end{enumerate}
 
Comparing the vertical scales in Figs.~\ref{fig:comparison-small}, we see that, for sufficiently large number of samples, the universal and extended asymptotic filters have a smaller one step prediction error than the empirical filter, but for $\lesssim 30$ samples, the empirical filter has a smaller prediction error. This reflects an initial transient, and is similar to the behavior in 
Fig.~\ref{fig:comparison-grand} where we compared the performance of the filters on synthetic data generated by numerically simulating the process in~\eqref{eq:continuum}. 

The error for the universal filter (Fig.~\ref{fig:comparison-small} (b)) is monotonically decreasing in $n$, the number of samples, roughly as a power law. In contrast, the error has interesting temporal structure for the empirical filter (Fig.~\ref{fig:comparison-small} (a)) and the extended asymptotic filter (Fig.~\ref{fig:comparison-small} (c)). Since we are plotting the absolute value of the error on a logarithmic scale, the downward spikes in the error are signatures of the times where the error changes sign. 

For the empirical filter, the error decreases until it hits a ``floor", roughly between $10^{-9}$ and $10^{-8}$. This floor is {\em independent of the time series and the sampling rate}, so it is inherent to the filter. In contrast, the error for the extended asymptotic filter has spikes are fixed times  $t = n\tau$, independent of the sampling interval $\tau$, so these spikes reflect features in the signal, and not the structure of the filter. Indeed, these spikes {\em correlate well} with the critical times $T_{crit}$ beyond which the given time series {\em cannot be realized in any solution} of~\eqref{eq:continuum}. Beyond the spikes, the one-step prediction error for the extended asymptotic filter increases, in contrast to the other two filters.
These observations support the following conclusions:
\begin{enumerate}
\item The universal filter has very small error as $n \to \infty$, but is not very discriminating. It tracks all functions with slowly decaying correlations, whether or not they come from solutions of~\eqref{eq:continuum}.
\item In tracking solutions of~\eqref{eq:continuum}, the empirical linear filter is very discriminating/nearly optimal among all linear filters with fixed coefficients and $L$ (a given number of) taps. However, it has a floor for its error reflecting the fact that we have ignored the $L+1$-th eigenvalue of the underlying sloppy model in the construction of this filter. Because of  the level repulsion between the eigenvalues of matrices from sloppy models~\cite{Waterfall_Sloppy_2006} the resulting error floor is pretty small when $L$ is moderate.
\item The extended asymptotic filter is (essentially) time varying so it has additional freedom which can be exploited to make filters that are both discriminating, and circumvent the above argument for the error floor.
\end{enumerate}
The numerical evidence for this picture is strong motivation to try and  formalize this intuition into rigorous mathematical statement and proofs.

\section{Discussion} \label{sec:discuss}

In this work, we have developed nonlinear, stochastic, reduced models for the evaporation process~\eqref{eq:continuum}, both empirically, i.e. in a data driven manner, and reductively from first principles, using the Mori-Zwanzig projection operator formalism, and by solving the appropriate Yule-Walker equations. The underlying system~\eqref{eq:continuum} has slow relaxation, and long memory, so it is of interest to see what intuition can be gleaned from our results, that might be generally applicable to other non-equilibrium systems with slow relaxation. 

Hamiltonian dynamics naturally supports an invariant measure on phase space, and the corresponding Liouvillian is {\em skew-symmetric} on the $L^2$ space for this invariant measure.  In contrast, our system~\eqref{eq:continuum} has no physically relevant invariant measure, is not the Liouvillian for any dynamical system, and is {\em symmetric} rather than skew-symmetric on an appropriate $L^2$ space. However, the evaporation process~\eqref{eq:continuum} is representative of systems with long memory and slow decay of correlations. So, we believe that our results do give some intuition for this class of non-equilibrium systems.

It is very surprising that a direct application of the Mori-Zwanzig projection operator formalism yields poor results for this system, considering the fact that the projection operator formalism is {\em exact} and is thus a natural starting point to make approximations. On the other hand, the MZ projection operator formalism, because it is exact, is constrained in ways that an empirical approximation is not. In this sense, a MZ decomposition, as in~\eqref{mz2}, has lots of parameters (the memory kernel $h_k$) and thus has the potential to overfit the training data, i.e. the sequence $\mathbb{E}[M_n]$ given by~\eqref{open-loop}. As a consequence of this overfitting, the resulting model has poor predictive power. This  argument suggests that the projection operator formalism is perhaps best suited for systems with an exponential decay of correlations, but perhaps not so well suited for systems with long memory \cite{Kawasaki_Theoretical_2000}.

Another lesson, that is reinforced by our results for the model process~\eqref{eq:continuum}, is the importance of picking the right ansatz and the right parameters for data-driven model reduction. In particular, both the MSM approach \cite{Chekroun_Predicting_2011,Kondrashov_Data_2015} and the NARMAX approach \cite{Chorin_Discrete_2015,Lu_Data_2017} to model identification use power series, and they assume that the higher order terms are smaller than the lower order ones. In our example, the model is homogeneous,  so these approaches will suggest that $M_{n}$ should be approximated as a linear combination of $M_{n-1},\ldots,M_{n-L}$. However, our results in sections~\ref{sec:linear}~and~\ref{sec:nonlinear} indicate that, for situations with slow relaxation,  one could do better by taking more general nonlinear homogeneous combinations, e.g. harmonic averages. 

 A key contribution of this work is development of the notions of {\em universal} and {\em asymptotic} filters. Indeed, the existence of these filters corroborates with theoretical ideas on {\em universality} for systems with slow decay of correlations \cite{Bouchaud_Aging_1999} and experimental observation of universality  in the glass transition \cite{Dixon_Scaling_1990}. These filters are also applicable in practice. Absent any other information, the universal filter~\eqref{state-space} is the optimal first order predictor \cite{Chorin_Optimal_2002} for systems with slow relaxation. However, this filter is unstable, so for particular problems, one has to go beyond this universal filter and develop asymptotic filters (e.g.~\eqref{asymp-filter}~and~\eqref{eaf}), i.e. stable, necessarily non-autonomous filters given by solutions of the appropriate Yule-Waler equations. These filters are asymptotic in that they converge to the universal filter as $n \to \infty$. We can view the asymptotic filters as analogs of the $t$-damping equation in \cite{Chorin_Optimal_2002}, albeit for our case where the memory is not short.

The extended asymptotic filter given by~\eqref{eaf} is a practical computational tool. It allows us to identify a single quantity $\kappa_n$ which accounts for the dependence of the evolution of macroscopic quantity $M_n$ on the microscopic state $\ket{\rho_n}$. For this reason, we will interpret $\kappa_n$ as the {\em microscopic age} of an oil distribution $\rho_n(w)$. This interpretation of $\kappa$ naturally follows from
\begin{align*}
M_{n} & = \int e^{-n w \tau} \rho_0(w) dw, \nonumber \\
\frac{M_{n-1}}{\kappa_{n-1}} \approx M_{n-1}-M_{n} & = \int (1-e^{-w \tau}) e^{-(n-1)w \tau} \rho_0(w) dw. \nonumber 
\end{align*}
Thus $\kappa_n$ is determined by the microscopic density distribution $\rho_n$ through
\begin{equation}
\label{micro-age}
\kappa_n \approx \frac{\int \rho_n(w) dw}{\int (1-e^{-w \tau}) \rho_n(w) dx}
\end{equation}
This relation can now be used to develop computationally efficient reduced models for the weathering of oil. For example, \eqref{eq:continuum} models a discrete release where oil is initially released at time $t=0$, and no further oil is added to the spill. In this situation, we can estimate $\kappa_n$ to obtain $\kappa_n \approx n$ which is indeed our motivation for defining the dynamical variable $\kappa_n$ in~\eqref{extended}. We can also consider the case of a continuous spill, in which case we have the microscopic model ({\em cf.} \eqref{Xfr-Koopman})
$$
\rho_{n}(w) = \Lambda \rho_{n}(w) + u_{n}(w)
$$
where $u_n$ represents the (deterministic or random) distribution of the oil added between time $n-1$ and $n$. Using the relation~\eqref{micro-age}, we can thus obtain effective equations for $\kappa_n$ to describe a continuous spill. This equation will replace the middle equation in~\eqref{extended}. We can now use the asymptotic filter coefficients in~\eqref{asymp-filter} along with the model definition~\eqref{extended} to obtain reduced stochastic models for a variety of oil spill scenarios. We will discuss these methods elsewhere.

As we have discussed previously, we can interpret our methods as a computational framework to study the evolution of an autonomous, linear, high-dimensional system (Eq.~\eqref{eq:species}) in the sloppy-model universality class \cite{Waterfall_Sloppy_2006}. There has been considerable work on understanding the geometry of the parameter landscape in sloppy models, and in identifying robust combinations of parameters that can be extracted from data \cite{Transtrum_Geometry_2011}. In contrast to earlier work, which is in a {\em static framework} (for an autonomous sloppy model, one attempts to model the system with another autonomous model, albeit one with robust parameters),  our work suggests that the robust parameters (in our case the filter coefficients $h^{(n)}_k$) might themselves evolve, even when the underlying model is autonomous. We hope to explore the consequences of this idea for other complex/nonlinear sloppy models.
 
Another direction we intend to pursue is to develop these ideas in a {\em non-parametric} setting \cite{Berry_Forecasting_2016}, as applied to spatially extended systems with slow relaxation. In particular, we want to combine our techniques for slow relaxation with the Nonlinear Laplacian spectral analysis (NLSA) method \cite{Giannakis_Nonlinear_2012,Comeau_Data_2015} which combines ideas from singular spectral analysis (see \cite{Vautard_Singular_1989}, \cite{Vautard_Singular_1992}) with the ideas of Coifman and Lafon \cite{Coifman_Diffusion_2006}  who introduced 
Diffusion maps (Laplacian eigenmaps). Diffusion maps  can be thought of as a powerful generalization of the Takens delay-coordinate embedding \cite{takens} to a form that is applicable to extended and high dimensional systems. The NLSA methods apply to large scale problems, but require short memory. Conversely, our methods in this work generates low dimensional reduced models from spatially homogeneous systems, but can handle long memory. 
A deeper comparison between these methods might reveal ways in which both of these techniques could be improved and made practical on large scale problems, such as those that arise from processes that are modeled by evolution equations with spatial dependence.

\begin{acknowledgements}
SV would like to acknowledge the many, very illuminating discussions with Kevin Lin who was very generous with his time and his ideas. We  are grateful to an anonymous referee for pointing out the potential connections between our work and the sloppy models universality class. This viewpoint turns out to be particularly fruitful.
\end{acknowledgements}

\appendix

\section{The Memory Kernel for Multiple Observables} \label{sec:multi}

One other comment is that we can indeed compute the memory kernel explicitly for the evaporation process \eqref{eq:continuum}, not just for the case with one observable, the mass $M_n$, but also more generally if we have a vector-valued linear observable $\Phi$, i.e $l$ scalar-valued observables $\Phi = \{\phi^1,\phi^2,\ldots,\phi^l\}^T$. Each scalar linear observable $\phi^i$ is given by an element of $\mathcal{H}^*$, and we will denote the corresponding bra-vector  by $\bra{\phi^i}$. Using the Gram-Schmidt procedure if necessary, we can assume that the vectors $\bra{\phi^i}$ given an orthonormal basis for their span, a $l$-dimensional subspace of $\mathcal{H}^*$. The orthogonal projection $P^* : \mathcal{H}^* \to \mathcal{H}^*$ onto this subspace is given by 
$$
\mathcal{P}^* = \ket{\phi^1} \bra{\phi^1} + \ket{\phi^2} \bra{\phi^2} + \cdots + \ket{\phi^2} \bra{\phi^2}.
 $$
 It follows that $\bra{\phi^i} P^* =  \bra{\phi^i} P = \bra{\phi^i}$ and~\eqref{master-eqn} gives
 $$
 \bra{\phi^i}\ket{\xi_{n+1}} = \sum_{k=0}^n \sum_{j=1}^l \bra{\phi^i} \Lambda (Q \Lambda)^k\ket{\phi^j} \bra{\phi^j}\ket{\xi_n} + \bra{\phi^i} (\Lambda Q)^{n+1} \ket{\rho_0}.
 $$
 The quantities $ \bra{\phi^i}\ket{\xi_{n+1}}$ are the entries of the ``vector" observable $\Phi_{n+1}$. Defining the matrices $H_k$ by $(H_k)_{ij} = \bra{\phi^i} \Lambda (Q \Lambda)^k\ket{\phi^j}$ for $k= 0,1,2,\ldots$ and  the (column) vectors $\beta_n$ by the entries $\beta^i_n  = \bra{\phi^i} (\Lambda Q)^{n+1} \ket{\rho_0}$, we have the Mori-Zwanzig decomposition
\begin{equation}
\label{mz3}
\Phi_{n+1} = \sum_{k=0}^n H_k \Phi_{n-k} + \beta_n.
\end{equation}
If $\ket{\rho_0}$ is in the span of $\ket{\phi^i}$, then $Q \ket{\rho_0} = 0$ so that the noise $\beta_n$ is identically zero. Taking $\ket{\rho_0} = \ket{\phi^1},\ket{\phi^2},\ldots,\ket{\phi^l}$ in turn, and collecting the corresponding column vectors $\Phi_n$ into a $l \times l$ matrix $\Xi_n$, we have
$$
(\Xi_n)_{ij} = \bra{\phi^i} \Lambda^n \ket{\phi^j} = \int_0^1 \phi^i(w) e^{-n w \tau} \phi^j(w) dw
$$
is a symmetric matrix for each $n$, and
$$
\Xi_{n+1} = \sum_{k=0}^n H_k \Xi_{n-k}, \qquad \mbox{ for } n = 0,1,2,\ldots
$$ 
 As before, we can determine the memory kernel $H_k$ using the $\mathcal{Z}$-transform. Defining the matrices
$$
\hat{\Xi}(z) = \sum_{n=0}^\infty z^{-n} \Xi_n, \qquad \hat{H}(z) = \sum_{n=0}^\infty z^{-n} H_n
$$
we get
$$
\hat{H}(z) = z(I - \hat{\Xi}(z)^{-1}).
$$
The matrix $\Xi_n$ is symmetric for all $n$, so that $H_n$ is also symmetric for all $n$. We expect that the norm $\|\Xi_n\|$ typically decays no faster than $1/n$. This is true for instance if the constant functions are in the range of $P$, or more generally if there are continuous functions $\psi$ with $\psi(0) > 0$ in the range of $P$. In this case, we expect that the norm of $H_n$ decays no faster that $1/(n \log^2(n))$ indicating again that, generically, one expects fat tails in the memory kernel for the system~\eqref{eq:continuum} if we use the Mori-Zwanzig decomposition based on any finite set of linear observables. 

\section{Orthogonal Dynamics} \label{sec:orthogonal}

We will now compute the statistics of the noise process $\beta_n$ in the Mori-Zwanzig decomposition $\eqref{mz2}$ with the usual approach through the study of the projection equation~\eqref{adjoint-predict} and the orthogonal dynamics~\eqref{orthogonal}. Since the orthogonal dynamics are linear, it suffices to solve the system 
$$
\bra{F_0} = \bra{\delta(w-x)} Q, \quad \bra{F_{n+1}} = \bra{F_n} \Lambda Q,\ \  n = 0,1,2,\ldots
$$
where $x \in [0,1]$ is fixed. A calculation reveals that, for any continuous function $\phi$,
$$
\braket{F_0}{\phi} = \braket{\delta(w-x)}{\phi} - \bra{\delta(w-x)} P \ket{\phi} = \phi(x) - \int_0^1 \phi(w) dw.
$$
We will thus associate $\bra{F_0}$ with the ``function" $F_0(w) = \delta(w-x) - 1$. We can follow this computation to solve the orthogonal dynamics equations recursively. For example, 
\begin{align*}
\braket{F_1}{\phi} & = \bra{\delta(w-x)-1} \Lambda \ket{\phi} - \bra{\delta(w-x)-1} \Lambda P \ket{\phi} \\
& = \int_0^1 (\delta(w-x)-1) e^{-w \tau} \phi(w) dw - \int_0^1 (\delta(w-x)-1) e^{-w \tau} dw \int_0^1 \phi(w) dw \\
& = \int_0^1 \left[e^{-x\tau} \delta(w-x)- e^{-w \tau} - e^{-x \tau} + \frac{1-e^{-\tau}}{\tau}\right] \phi(w) dw,
\end{align*}
so that $\bra{F_1}$ corresponds to the function $F_1(w) = e^{-x\tau} \delta(w-x)- e^{-w \tau} - e^{-x \tau} + \frac{1-e^{-\tau}}{\tau}$. Using the fact that $Q$ and $\Lambda$ are self-adjoint operators on $\mathcal{H}$, and further $\bra{\psi} \Lambda \ket{\phi} = \int \psi(w) e^{-w \tau} \phi(w) dw$ so that $\Lambda$ is diagonal on the ``basis" $\{\delta(w-x)\}_{\{x \in [0,1]\}}$, an inductive argument shows that $F_n(w) = e^{-nx\tau} \delta(w-x) + \Psi_n(w;x)$ where $\Psi_n$ is a smooth, symmetric function $\Psi_n(w;x) = \Psi_n(x;w)$. We will use these conclusions to verify the full solution for $\bra{F_n}$ that we obtain below by independent means.

Consider the $\mathcal{Z}$-transform $\hat{\bra{F}} = \sum z^{-n} \bra{F_n}$. The orthogonal dynamics imply 
$$
\hat{\bra{F}} (1 - z^{-1} \Lambda Q) = \hat{\bra{F}} - z^{-1} \hat{\bra{F}} \Lambda  +  z^{-1} \hat{\bra{F}} \Lambda \ket{1}\bra{1} = \bra{F_0}.
$$   
Using the ansatz $\hat{F}(z,x,w) = \hat{A}(z,x) \delta(w-x) + \hat{\Psi}(z,x,w)$ corresponding to a decomposition of $F_n$ into its singular and regular parts, we get the pair of equations
\begin{align*}
(1- z^{-1} e^{-x \tau}) \hat{A} & = 1, \\
\hat{\Psi} - z^{-1} e^{-x \tau} \hat{\Psi} + z^{-1} e^{-w \tau} \hat{A} + z^{-1} \int e^{-x \tau} \hat{\Psi} dx & = -1,
\end{align*}
where we have suppressed the arguments $(z,x,w)$ for $\hat{A}$ and $\hat{\Psi}$ for clarity. We can solve the first equation to obtain
$$
\hat{A} = \frac{1}{1-z^{-1} e^{-x \tau}}.
$$
Using this in the second equation, we obtain
$$
\hat{\Psi} = -\frac{1}{(1-z^{-1} e^{-x \tau}) (1-z^{-1} e^{-w \tau})} - \frac{z^{-1} C(z,w)}{1-z^{-1} e^{-x \tau}},
$$
where $C(z,w) = \int e^{-x \tau} \hat{\Psi} dx$ is determined in terms of the required solution $\hat{\Psi}$ self-consistently. Multiplying by $e^{-x \tau}$ and integrating in $x$, and solving the resulting equation for $C(z,w)$, we obtain
$$
C(z,w) = -\frac{z\left(\tau -\log \left(1-e^{\tau } z\right)+\log
   (1-z)\right)}{\left(1-z^{-1} e^{-w \tau}\right) \left(\log (1-z)-\log \left(1-e^{\tau }
   z\right)\right)}.
$$
Using this result in the computation for $\hat{\Psi}$ gives
$$
\hat{\Psi}(z,x,w) = \frac{\tau }{\left(1-z^{-1}e^{-w\tau}\right) \left(1-z^{-1}e^{-x \tau}\right) \left(\log (1-z)-\log \left(1-e^{\tau } z\right)\right)}.
   $$
This gives a complete solution of orthogonal dynamics equation by 
$$
\hat{F}(z) =  \frac{\delta(w-x)}{1-z^{-1} e^{-x \tau}} + \frac{\tau }{\left(1-z^{-1}e^{-w\tau}\right) \left(1-z^{-1}e^{-x \tau}\right) \left(\log (1-z)-\log \left(1-e^{\tau } z\right)\right)}.
$$
The singular part of $F_n$ is therefore $e^{-n \tau x} \delta(w-x)$ as we noted above. Further, the regular part $\hat{\Psi}$ is symmetric in $w$ and $x$, implying this property for each of the functions $\Psi_n$. Finally, for an observable given by a continuous function $g$, the solution to the orthogonal dynamics is given by
$$
\sum_{n=0}^\infty z^{-n} \mel{g}{Q (\Lambda Q)^n }{\phi} = \int_0^1 \int_0^1 g(x) \hat{F}(z,w,x) \phi(w) dw dx.
$$
For the observable $M_{n}$, the prediction for the total mass at the next time step, we have $\bra{g} = \bra{1} \Lambda$. The $\mathcal{Z}$-transform of the memory kernel is given by
\begin{align*}
H(z) = \sum_{k \geq 1} z^{-k} h_k & = z^{-2} \sum_{k \geq 1} z^{-k+2} \mel{1}{(\Lambda Q)^{k-1} \Lambda}{1} \\
& = z^{-1} \expval{\Lambda}{1}+z^{-2}\sum_{n \geq 0} z^{-n}  \mel{1}{\Lambda Q (\Lambda Q)^n \Lambda }{1} \\
& = z^{-1} \frac{1-e^{-\tau}}{\tau} + z^{-2}  \int_0^1 \int_0^1 e^{-x \tau} \hat{F}(z,w,x) e^{-w \tau} dw dx \\
& = \left[1 - \frac{\tau}{\log(e^\tau z -1) - \log(z-1)}\right].
\end{align*}
The $\mathcal{Z}$-transform of the expected values of the noise sequence $\beta_n$ is given by 
$$
\sum z^{-n} \mathbb{E}[\beta_n] = \sum z^{-n} \mathbb{E}[\braket{F_n}{\rho_0}] =  \int_0^1 \int_0^1 e^{-x \tau} \hat{F}(z,w,x) dw dx = 0 
$$
and the correlations between the noise $\beta_n$ and the mass $M_j$ are given by $\mathbb{E}[\beta_n M_j]  = \bar{\sigma}^2 \mel{1}{(\Lambda Q)^n \Lambda^j}{1}$. (See section~\ref{sec:analysis}). Taking the (two index) $\mathcal{Z}$-transform, noting that $\beta_0 = 0$,  we have
\begin{align*}
\sum_{n \geq 1} \sum_{j \geq 0}  z^{-n} \zeta^{-j} \mathbb{E}[\beta_n M_j] & = \bar{\sigma}^2\sum_{n \geq 1} \sum_{j \geq 0} z^{-n} \zeta^{-j} \mel{1}{(\Lambda Q)^{n} \Lambda^j}{1} \\
& = \bar{\sigma}^2z^{-1}\sum_{n \geq 0} \sum_{j \geq 0} z^{-n} \zeta^{-j}  \mel{1}{\Lambda Q (\Lambda Q)^n \Lambda^j }{1} \\
& = \bar{\sigma}^2z^{-1} \int_0^1 \int_0^1 \frac{e^{-x \tau} \hat{F}(z,w,x) }{1-\zeta^{-1} e^{-w \tau}} dw dx \\
& = \frac{\bar{\sigma}^2}{\tau} z^{-1} \Bigg[\frac{1-z^{-1}e^{ -\tau}}{(1-z^{-1})(1-z^{-1}e^{-\tau})} \\ 
& + \frac{\log \left(\frac{1-z^{-1}}{1-z^{-1}e^{-\tau}}\right) \left( \log \left(z \log \left(\frac{\zeta -1}{\zeta  e^{\tau}-1}\right)- \zeta \log \left(\frac{z -1}{z  e^{\tau}-1}\right) \right) \right)}
   {(z-\zeta ) \log \left(\frac{z-1}{e^{\tau } z-1} \right)}\Bigg].
\end{align*}
It is not true that $\mathbb{E}[\beta_n M_j] = 0$ if $n > j$, as one would expect in the Mori-Zwanzig decomposition for a system with an invariant measure. In particular, 
$$
\mathbb{E}[\beta_2 M_1] = \frac{\bar{\sigma}^2 (1-e^{-\tau})((\tau-2) +(\tau+2)e^{-\tau})}{2\tau^2} \neq 0
$$

\section {Sampling Initial Conditions}
\label{apndx:weak-limit}

For any prescribed value $0 < \bar{\sigma}^2 < \infty$, we can indeed find a family of $I$-dependent distributions such that 
$$
\mu_\gamma \to 1,  \frac{\sigma_\gamma^2}{I} \to \bar{\sigma}^2 \mbox{ as } I \to \infty 
$$ 
by appropriately truncating and rescaling a distribution that has finite mean but infinite variance. For example, the function
$$
f(x) = \begin{cases} \frac{9}{10} & 0 \leq x \leq \frac{2}{3}, \\ \frac{9}{10}\left(\frac{3x}{2}\right)^{5/2} & x > \frac{2}{3} \end{cases}
$$
satisfies $f \geq 0$ on $(0,\infty)$ and $\int_0^\infty f(x) dx = 1$, so $f$ is indeed a nonmalized density on $(0,\infty)$. Further $\int_0^\infty x f(x) dx = 1$ and $\int_0^L x^2 f(x) dx \sim \sqrt{\frac{32}{75} L}$ for $L \gg 1$. We can therefore define a sequence of $I$ dependent distributions by truncating the support of $f$ and renormalizing to have unit mass, i.e. 
$$
f_I(x) = \begin{cases} c_I f(x) & 0 \leq x \leq L_I \\ 0 & x > L_I, \end{cases}
$$
where $L_I$ is any sequence satisfying $L_I \geq 2/3$ for all $I$, $L_I \nearrow \infty$  and $\displaystyle{\sqrt{\frac{32}{75 I^2} L_I} \to \bar{\sigma}^2}$ as $I \to \infty$. Given such a sequence $L_I$, the normalization $c_I$ is determined by $\int_0^I f_I(x) dx = 1$ so that $c_I \to 1$.

\section{Asymptotic solutions of the Yule-Walker equations} \label{sec:cauchy}

We seek a solution to~\eqref{eq:asymp-matrix} as an asymptotic series in $n$, i.e. solutions of the form 
\begin{equation}
h^{(n)}_j = a^0_j + \frac{1}{n} a_1^j + \frac{1}{n^2} a_2^j + \cdots .
\label{eq:asymp-expnsn}
\end{equation}
The difficulty in solving this system is evident if we expand the coefficient matrix $A$ as a power series in $n$:
$$
A = \frac{1}{2n}\begin{pmatrix} 1 & 1 & \cdots &  1 \\
 1 & 1 & \cdots &  1 \\
\vdots & \vdots & \ddots & \vdots \\
1 & 1 & \cdots &  1 \end{pmatrix}
+
\frac{1}{4n^2}\begin{pmatrix} 2 & 3 & \cdots &  L+1\\
 3 & 4 & \cdots &  L +2\\
\vdots & \vdots & \ddots & \vdots \\
L +1& L+2 & \cdots &  2L\end{pmatrix}
+ \cdots .
$$
Assuming $L \geq 3$, the two matrices displayed in the expansion of $A$ are singular. The first matrix has rank 1, the second has rank 2. Indeed the first $L-1$ matrices in the expansion of $A$ are all singular and their (row) nullspaces are nested 
$$
v^T \begin{pmatrix} 2 & 3 & \cdots &  L+1 \\
 3 & 4 & \cdots &  L+2 \\
\vdots & \vdots & \ddots & \vdots \\
L +1  & L+2 & \cdots &  2L \end{pmatrix} = 0 \implies v^T \begin{pmatrix} 1 & 1 & \cdots &  1 \\
 1 & 1 & \cdots &  1 \\
\vdots & \vdots & \ddots & \vdots \\
1 & 1 & \cdots &  1 \end{pmatrix}
=0,$$
and so on. The determinant of $A$ is thus very close to zero ($\det{A} \sim O(n^{-L^2})$ as we see below) so it is not  clear that we have solutions for $h^{(n)}$ where the leading order behavior stays $O(1)$ instead of diverging with $n$. Proving the boundedness of $h^{(n)}$ and determining the $O(1)$ solution thus requires consideration of $L$ {\em solvability conditions} given by the vectors that span the common (row)-nullspaces of the initial $j$ terms in the expansion of $A$ for $j=1,2,\ldots,L-1$. Higher order terms will require even longer expansion of the matrices and more solvability conditions. 

In the general case of a process with slowly decaying correlations, it is still true that the matrix of coefficients in the Yule-Walker equation is nearly singular, and one does have to go through the process described above to find optimal, reduced dimensional, models for such systems. For the evaporation process \eqref{eq:continuum} however, the coefficient matrix has a special structure, that we exploit to find the solutions for the optimal filter $h^{(n)}$. The matrix $A$ is a {\em Cauchy matrix} \cite{Polya_Szego_II} i.e its entries are of the form $A_{ij} = 1/(x_i-y_j)$. In particular, we can choose $x_i = 2n-i$ and $y_j = j$. The determinant of a Cauchy matrix $A_{ij} = 1/(x_i-y_j)$ is given by \cite{Polya_Szego_II}
$$
\det{A} = \frac{\prod_{i > j} (x_i-x_j)(y_j-y_i)}{\prod_i \prod_j (x_i-y_j)}.
$$
For the particular matrix $A$ from above, the terms in the numerator are all bounded by $L$ and the terms in the denominator are all $\approx 2n$ if $n \gg L$. Consequently, 
$\det{A}  \sim O(n^{-L^2})$. The matrix $\hat{A}_m$ obtained by replacing  the $m$-th column of $A$ by the vector $v_i = \frac{1}{2n -i}$ is also a Cauchy matrix $\hat{A}_{ij} = 1/(x_i - \hat{y}_j)$, with the same choice $x_i = 2n-i$ and 
$$
\hat{y}_j  = \begin{cases} y_j & j \neq m, \\ 0 & j = m. \end{cases}
$$ 
Cramer's rule now yields, 
\begin{equation}
h^{(n)}_m = \frac{\det{\hat{A}}}{\det{A}} = \prod_{i \neq m} \frac{i}{i-m} \prod_i \frac{2n - i -m}{2n - i}.
\end{equation}

\bibliographystyle{siam}      
\bibliography{mz-aging,aging-oil}   

\end{document}